\RequirePackage{ifpdf}
\documentclass[11pt,a4paper]{article}
\pdfoutput = 1
\usepackage{jcappub}
\usepackage{mathrsfs}
\usepackage{dcolumn}
\usepackage{bm}
\usepackage{color}
\usepackage{ulem}

\usepackage{multirow}

\providecommand{\beqa}{\begin{eqnarray}}
 
 \providecommand{\rm}{\mathrm}
\providecommand{\eeqa}{\end{eqnarray}}

\newcommand{\beq}{\begin{equation}}
\newcommand{\eeq}{\end{equation}}

\def\46{{\scriptscriptstyle {\rm 4-6}}}
\def\24{{\scriptscriptstyle {\rm 2-4}}}

\providecommand{\MM}{\mathbb{M}}

\def\Tr{  \mbox{Tr}   }

\title{Non-Minimal M-flation}

\author[a]{Amjad Ashoorioon,}
\author[a,b]{Kazem Rezazadeh}
\affiliation[a]{\small School of Physics, Institute for Research in Fundamental Sciences (IPM),\\
P.O. Box 19395-5531, Tehran, Iran}
\affiliation[b]{\small Department of Physics, University of Kurdistan, Pasdaran Street, P.O. Box 66177-15175, Sanandaj, Iran}

\emailAdd{amjad@ipm.ir}
\emailAdd{rezazadeh86@gmail.com}

\abstract{We show how in a matrix inflationary model in which there is a non-minimal coupling between the matrix inflatons and gravity --hence dubbed Non-$\MM$-flation-- some of the disadvantages of the minimal model can be avoided.  In particular, the number of D3 branes can be reduced substantially to $\lesssim \mathcal{O}(100)$, which can alleviate the ``potential'' backreaction problem of large number of D3 branes on the background geometry. This is achieved by values of non-minimal coupling of order few hundred, which is much smaller than that of Higgs Inflation. The prediction of the model in the symmetry breaking part of the potential, which is a local attractor and can support eternal inflation, is compatible with the latest PLANCK results. In contrast to the minimal model, the spectator fields can partially or completely reheat the universe, depending on the symmetry-breaking vacuum expectation value and the non-minimal coupling parameter. We also comment on how the presence of gauge species keep the UV cutoff at around the Planck scale in the Einstein frame and, in contrast to the Higgs inflation, the problem of field displacements beyond the cutoff does not occur.}

\subheader{IPM/P-2019/018}

\begin{document}
\maketitle

\section{Introduction}
\label{section:introduction}

Embedding the paradigm of inflation in the landscape of string theory, soon turned out to be a tedious task, despite how rich and vast the landscape looked like in the inception of its formulation. From two classes of small and large single field models, only the former \cite{Kachru:2003sx} was thought to be possible to formulate in the string theory until about a decade ago \cite{Baumann:2006cd}, before the devise of monodromy inflation \cite{Silverstein:2008sg,McAllister:2008hb}. In such single moduli models, the stabilization of volume modulus {\it generically}, couples the inflaton conformally to gravity, causing the notorious $\eta$-problem. This happens despite -- and in fact because of -- the large warping of the internal manifold, produced by internal fluxes. Before monodromy inflation, of course, invoking the assisted mechanism \cite{Liddle:1998jc}, collaborative enhancement of several moduli was invoked to enhance the Hubble friction and realize inflation from otherwise individually steep potentials \cite{Becker:2005sg, Ashoorioon:2006wc, Dimopoulos:2005ac}. These models can in general produce an observational B-mode signature with tensor-to-scalar ratio, $r$, larger than $0.01$ despite individual field displacements much less than Planck mass, which is often demanded for $r\gtrsim 0.01$ \cite{Lyth:1996im}. The Planck mass in N-flation \cite{Dimopoulos:2005ac} though, is sensitive to radiative corrections of the scalar moduli to the graviton propagator, which diminishes the UV cutoff in such theories \cite{Dvali:2007hz,Ashoorioon:2011ki} and the problem of sensitivity to field displacements beyond the cutoff resurfaces again. In multiple M5 brane inflation \cite{Becker:2005sg, Ashoorioon:2006wc}, it is impossible to sustain inflation for the enough number of e-folds required to solve the problems of the standard Big Bang cosmology \cite{Ashoorioon-unpublished}. Non-minimal couplings to gravity can also be generated through the scalar field loops \cite{Ashoorioon:2011aa} \footnote{For realizations of accelerating expansion related to late time universe within string and F-theories, please see \cite{Heckman:2018mxl, Heckman:2019dsj}}.

Matrix inflation could be regarded as the third avenue in the construction of inflation in string theory, where the other two approaches are open string \cite{Kachru:2003sx} and closed string inflation \cite{Kallosh:2007cc,Silverstein:2008sg}.  As suggested by its name, the model is driven by matrices, which correspond to the dimensions perpendicular to a stack of $N$ D-branes. The dimensions perpendicular to the stack of D-branes are scalars in the adjoint representation of $U(N)$ and hence they are $N\times N$ hermitian matrices. The ones that are parallel to the D-branes correspond to $U(N)$ gauge fields that are also matrices. The original realization \cite{Ashoorioon:2009wa} was motivated by the dynamics of a stack of $N$ D3 branes (or concentric stack of $N$ and $M$ D3 branes \cite{Ashoorioon:2009sr}) probing a specific background geometry which  was sourced by a specific background flux. In the prime picture, however, the $U(N)$ was taken to be global instead of local and the constraints that the background geometry and the flux would need to satisfy in order for them to be a solution to the supergravity equations of motion were ignored. That led to a landscape of inflationary models with quartic polynomial potentials, among which usual chaotic models, like $m^2\phi^2$ and $\lambda\phi^4$, or hilltop inflationary models \cite{Boubekeur:2005zm} or inflection-point inflation \cite{Allahverdi:2006iq} exist. Due to the specific form of the potential realized from the expansion of the DBI action for the system in $\alpha'$, it was assumed that three of the perpendicular dimensions are only assertive in the inflationary dynamics. This allowed to relate these three dimensions with the three $N\times N$ generators of the $SU(2)$ group algebra.  In the gauged model, which is motivated from string theory, we also assume that the background in ten dimension satisfies the supergravity equations of motion, which enforces the potential for the effective potential for the effective inflaton to take a displaced Higgs potential with super-Planckian vacuum expectation values (vev's). The $U(N)$ gauge group is assumed to be local in these Gauged M-flation picture \cite{Ashoorioon:2011ki, Ashoorioon:2014jja}.

In order to suppress the self-coupling of the chaotic inflationary models realized in this matrix setup, from bare couplings that one would naturally expect to be of order one, to the values  required to explain the observed amplitude of density perturbations, usually a large number of $D3$-branes, $N\sim 10^5$, is required. Such large number of D3 branes and the flux couples to it can backreact on the background geometry which is generated by the exposed flux. It would be appealing if one could somehow reduce the required number of D3-branes in the model. One way to do this is reducing the string coupling $g_{_{S}}$ to a very tiny values. This will however itself is a fine-tuning and in violation of the original purpose of the matrix structure, which aimed to suppress the couplings to the observed value, using the multiplicity of the D3 branes.

Another annoying issue with the original setup is that the configuration of the matrices that lead to inflation, namely the $SU(2)$ configuration, is not an attractor in the whole hill-top region which is still consistent with the latest PLANCK results \cite{Akrami:2018odb}. The region beyond the symmetry-breaking vacuum, which was an attractor for all values of the inflaton field predicts a value of tensor-to-scalar ratio $0.1\lesssim r\lesssim 0.2$, which was in the sweet spot of the BICEP2 \cite{Ade:2014xna}. However, soon it turned out the signal is mostly coming from foreground dust polarization rather than primordial quantum fluctuations \cite{Flauger:2014qra, Mortonson:2014bja, Adam:2014bub, Array:2015xqh}. The upper bound of PLANCK 2018 results set on the tensor-to-scalar ratio, certainly rules out this region of potential, assuming Bunch-Davies initial conditions\footnote{One can lower the tensor-to-scalar ratio in this region of potential, using super-excited initial states as in \cite{Ashoorioon:2013eia}, but to prepare these initial conditions, one would need to depart from Lorentzian dispersion relations at very high physical momenta for each mode  \cite{Ashoorioon:2017toq,Ashoorioon:2018uey}.}. The hilltop region in which inflaton ends in the symmetry-breaking vacuum, as we will see, predicts a finite number of e-folds $N_e\sim 100$, as one of the spectator modes become tachyonic for values of inflaton beyond the one at which the largest scale crosses the horizon. By itself this is no problem, as in order to solve the problems of the standard Big Bang cosmology we would need only 60 e-folds. However, in this region of potential it is not possible to realize eternal inflation, often provoked to populate the stringy landscape  \cite{Linde:1986fc}. Also the problem of classical initial condition for the field that may miss this segment of the potential becomes another challenge that one would have to deal with in absence of eternal inflation and the landscape picture. Nonetheless, finite number of e-folds may have interesting observational consequences \cite{Freivogel:2005vv}. In the part of hilltop potential in which the inflaton ends up in the symmetric vacuum, there are multitude of the spectator modes that can become tachyonic around the $SU(2)$ direction during inflation, which precludes the configuration to be even a local attractor during the 60 e-folds of inflation in this part of the potential.

In the matrix setup, perpendicular to the direction of the inflaton, there are many fields that are frozen classically, hence called {\it spectators}. Their mass is a function of the  inflaton and as it oscillates at the bottom of potential, the mechanism has the potential to produce particles non-adiabatically \cite{Abbott:1982hn, Traschen:1990sw, Shtanov:1994ce, Kofman:1994rk, Kofman:1997yn}. The couplings of the preheat fields to the inflaton are related to the inflaton's self-couplings which are fixed by the CMB observations. That would allow for the inflaton's energy at the end of inflation to be transferred to the spectators so that reheating occurs. On the other hand, as stated previously, some of the spectator modes become tachyonic during inflation and in fact all of them become tachyonic in a small region around the symmetric vacuum. However the region of the potential for which the inflaton ends in the symmetric vacuum is not a local attractor for the $SU(2)$ configurations. For inflation to work the vev of the symmetry-breaking vacuum has to be much bigger than the Planck mass, $M_P$. This will also prevent the inflaton from rolling over the hilltop region and oscillating around the symmetric vacuum to be able to take the advantage of the tachyonic spectator modes as preheat fields.

As explained above, inflationary $\eta$-problem has always been regarded as the Achilles heel of the inflationary setups realized in string theory. Regardless of the origin of the inflaton coupling with gravity and the value of non-minimal coupling, $\xi$, inflaton couplings to gravity through terms proportional to $\xi R\phi^2$ are expected to show up once one compactifies to four dimensions. In this paper, we would like to use this often-regarded  intimidating factor, to alleviate the problems involves with the matrix inflationary setup.

The structure of the paper is as follows. First we succinctly  review the setup of matrix inflation and the potential that arises from the  interaction of a stack of D-branes with a higher dimensional form flux. Then we elaborate in detail the shortcomings of  the minimal setup of matrix inflation. Then we elaborate how non-minimal couplings can arise in the inflationary setups realized within string theory. One mechanism is the loop corrections of the species to gravity which we show can at most create non-minimal couplings of order one. Such small corrections, although cannot address the large number of D-branes, which is required to suppress the couplings to tiny values from observation, can to some extent relieve the tension of the model with the PLANCK data and, as we will see in the next chapters, can partially transfer the energy from the $SU(2)$ sector inflaton to the spectator modes. For large values of the coupling of the inflaton to gravity, like the $\mathbb{K}$L$\MM$T setup \cite{Kachru:2003sx}, one may be able to summon the dependence of the superpotentials to the position of the D3 branes moduli. However, contrary to their case, in which this contribution is tuned to  cancel the conformal coupling of the inflaton to gravity, we consider the case that this contribution causes a non-minimal coupling much larger than one. We show that with large non-minimal couplings, one can mitigate all the aforementioned problems. We compute the predictions of the inflaton in the $n_{_{S}}-r$ plane. Contrary to the case of Higgs inflation \cite{Bezrukov:2007ep}, where the inflaton's vev and self coupling is fixed by the experiment, here we are left with the freedom in the choice of these parameters in the potential. Also in contrast with the case of minimal model, $\MM$-flation, the vev of the inflaton is no longer required to take super-Planckian values to conform to the demands of the CMB observations. We also compute the amplitude of isocurvature perturbations from various sectors numerically and show that they are mostly negligible at the end of inflation. In the next section, we address the preheating in non-$\MM$-flation and show that one can successfully deplete part or all of the energy of the  inflaton. In the last section of the article, we address the issue of UV cutoff in the model and argue that, contrary to the case of Higgs inflation  \cite{Bezrukov:2007ep}, the field displacements could be kept smaller than the UV cutoff of the model. We conclude the paper and provide directions for future research in the last part of the paper. The paper contain two appendices where in the first one we show that in the limit of large non-minimal couplings, $\xi\gg 1$, the predictions of all non-$\MM$-flationary models approach a single point in the $n_{_{S}}-r$ plane. In the second one we compare our exact numerical approach in finding the canonical field in the Higgs inflation setup and show that this exact approach imparts a correction of $10^{-4}$ to the predictions of the model. This could be important in light of future CMB experiments, which measure the quantities of interest with an unprecedented precision.

\section{$\mathbb{M}$-flation: a Review}

\subsection{Background Dynamics}

 The ingredients of the matrix inflation is a stack of $N$ string theory D3 branes in a type IIB supergravity background,
 \beq\label{sugra-background}%
ds^2 =-2dx^+dx^--\hat m^2 \sum_{i=1}^3 (x^i)^2 (dx^
+)^2+\sum_{I=1}^8d
x_I dx_I,
\eeq
sourced by an $RR$ six form flux, which could arise from a distribution of D5 branes,
\beq
C_{+123ij}= \frac{2\hat\kappa}{3} \epsilon_{ijk} x^k\,.
\eeq
$\hat{\kappa}$ parameterizes the strength of the $RR$  six form flux, $C_{+123ij}$, which has  two legs along the directions
transverse to the D3-branes. It was \textit{assumed} that the background geometry is described by \eqref{sugra-background} at an almost string length scale but could become a Ricci flat geometry that can become compactified on a $T^6$ or CY$_3$, which would then render the four dimensional Planck mass finite. It was also posited that the process of compactification to four dimension cause a \textit{minimal} coupling between the transverse dimensions of the stack of D3 branes and gravity. This brought action of minimal M-flation, or $\MM$-flation henceforth, to the form (please see \cite{Ashoorioon:2009wa, Ashoorioon:2011ki} for details),
\beq\label{action}
 S_{{}_{\rm \mathbb{M}-flation}}=\int d^{4} x \sqrt{-g} \left(\frac{-M_{P}^{2}}{2} R - \frac{1}{4} \Tr(F_{\mu\nu}F^{\mu\nu})- \frac{1}{2}
\sum_{i} \Tr  \left( D_{\mu} \Phi_{i} D^{\mu} \Phi_{i}
\right) - V(\Phi_{i}, [ \Phi_{i}, \Phi_{j}] ) \right) \, .%
\eeq %
We work in the units that the reduced Planck mass $M_{P}\equiv\left(8\pi G\right)^{-1/2}$, and also we assume the metric signature as $(-,+,+,+)$. We denote the matrices fields by $\Phi_i,\, i=1,2,3$. The matrices $\Phi_{i}$ are proportional to three out of six dimensions transverse to the D3-branes and the potential takes the form%
\beq\label{The-Potential}%
V= \Tr  \left( - \frac{\lambda}{4}  [ \Phi_{i},
\Phi_{j}] [ \Phi_{i}, \Phi_{j}] +\frac{i \kappa}{3} \epsilon_{jkl}
[\Phi_{k}, \Phi_{l} ] \Phi_{j} +  \frac{m^{2}}{2}  \Phi_{i}^{2}
\right),%
\eeq%
where, as emphasized before, $i$ runs from $1\ldots 3$. The quadratic and cubic couplings, $\lambda$ \& $\kappa$  respectively, are related to the string coupling
and the strength of the Ramond-Ramond antisymmetric form, and $m$ is
the same $\hat m$ that appears in the metric:
\beq%
\lambda=8\pi g_{_{S}}=2g^2_{{}_{YM}}\ ,\qquad \kappa= \hat{\kappa} g_{_{S}}\sqrt{8\pi g_{_{S}}}\ ,\qquad m^2=\hat m^2 . %
\eeq%
In order for the background solution \eqref{sugra-background} to be a solution to
the supergravity equation of motion with a constant, the following relation between the parameters $\lambda$, $m$ and $\kappa$ should hold,
\beq\label{susy-cond}%
\lambda m^2=4\kappa^2/9\,. %
\eeq %

As discussed in \cite{Ashoorioon:2009wa,Ashoorioon:2009sr,Ashoorioon:2011ki, Ashoorioon:2014jja},  one can simplify the background dynamics by identifying the three $N\times N$ scalar matrices, which would contain $3N^2$ degrees of freedom, to be proportional to the $N$ dimensional generators of the $SU(2)$ algebra, $J_{i}$, with a single proportionality factor $\hat \phi$,
\beq \label{phiJ}%
\Phi_{i} =
\hat \phi(t) J_{i}\ , \quad \quad i=1,2,3,%
\eeq%
 $\Phi_{i}$ and $J_{i}$ are hermitian and hence $\hat\phi$ is real scalar field. It is easy to see that one can consistently turn off the
gauge fields $A_\mu$ in the background, and hence, the classical
inflationary trajectory takes place in the scalar fields $\Phi_i$ sector.

Plugging the ansatz  \eqref{phiJ} into the action (\ref{action}) one obtains
\beq
S= \int
d^{4} x \sqrt {-g} \left[- \frac{M_{P}^{2}}{2} R+ \Tr J^{2}  \left( -
\frac{1}{2}  \partial_{\mu} \hat \phi  \partial^{\mu} \hat \phi
-\frac{\lambda}{2}  \hat \phi^{4}  + \frac{2 \kappa}{3} \hat
\phi^{3} - \frac{m^{2}}{2} \hat \phi^{2} \right) \right] \, , \eeq
where $\Tr J^{2} = \sum_{i=1}^{3} \Tr (J_{i}^{2})   = N(N^{2}
-1)/4$.
Upon the field redefinition%
\beq \label{phi-scaling}%
 \hat \phi = \left(  \Tr
J^{2}   \right)^{-1/2} \phi = \left[ \frac{N}{4}(N^{2}-1)
\right]^{-1/2} \, \phi \, , %
\eeq%
one can make the kinetic energy for the new field $\phi$ canonical, while the potential takes the form,
\beq\label{Vphi}%
V_0(\phi)= \frac{\lambda_{eff}}{4} \phi^{4} -
\frac{2\kappa_{eff}}{3} \phi^{3} + \frac{m^{2}}{2} \phi^{2}= \frac{\lambda_{eff}}{4} \phi^{2} (\phi-\mu)^2 \, , %
\eeq%
where%
\beqa \label{lameff}%
\lambda_{\rm eff} = \frac{2 \lambda}{\Tr J^{2}} = \frac{8 \lambda}{ N
(N^{2}-1)}  \ , &&\quad \kappa_{\rm eff} = \frac{ \kappa}{\sqrt{\Tr J^{2}}
} = \frac{2
\kappa}{\sqrt{N(N^{2}-1)}}\,,\\  &&\mu=\frac{\sqrt{2}m}{\lambda_{\rm eff}}\, . %
\eeqa%
The effective inflaton has a displaced Higgs-like potential with super-Planckian vev's. It is easy to show that  for values of undressed couplings of order one, in order to satisfy the constraints from the CMB, one needs $N\sim {\rm few}\times 10^4-10^5$ D3 branes, depending on whether inflation happens in the hilltop region, $0<\phi<\mu$, or in the symmetry breaking region, $\phi>\mu$. The typical physical field displacement, $\Delta\hat\phi$, is around $10^{-6} M_{P}$ which is much smaller than the Planck mass.


The individual potential, before the canonicalization of the kinetic term, satisfies the de-Sitter criterion of the  conjecture. With  displacement much less than $M_{P}$, and with $\lambda, ~\kappa\sim \mathcal{O}(1)$, the relative slope of the potential, $|\nabla V|/V\sim \mathcal{O}(1)$ before one makes the kinetic term canonical. In fact, the $SU(2)$ sector ansatz induces a nontrivial field space metric similar to the approach of \cite{Achucarro:2018vey}, which facilitates inflation on potentials with large slope by introducing turn in the trajectory.

\subsection{Spectrum of Spectators}

In the $SU(2)$ sector, the classical dynamics of the system is reduced to a single scalar field, $\hat \phi$. However noting the matrix nature of the ingredients, the physical number of degrees of freedom (dof) is much larger. These dof's are even though frozen classically, for which we call them spectators, can still fluctuate quantum mechanically and have significant thumbprints during or after inflation, {\it i.e.} during preheating. Some of these observational signatures have been spelled out in previous works on $\MM$-flation, as isocurvature spectra or high frequency gravitational waves from inflation \cite{Ashoorioon:2009wa, Ashoorioon:2014jja, Ashoorioon:2013oha}. Below we will just briefly mention the categories and masses of these spectators. In the derivation of the mass of these spectators, one has to perturb the field around its background value and diagonalize the resulted mass operator generated in second order in perturbations. In the bosonic matrix inflationary model\footnote{In the full supersymmetric model, besides the scalar and gauge fields, there will be fermionic fields. We postpone the investigation of these fermionic spectators to a future investigation.}, depending on the sector that  these spectators stem from, one can group the spectators to two categories:

\paragraph{\small{Scalar Spectators Modes}}

As the name suggests, these are the {\it physical} modes that originate from the scalar fields, $\Phi_i$, and are perpendicular to the $SU(2)$ sector, $\Phi_i=\hat\phi J_i$. Depending on the eigenvalue, there are two distinct categories:

\begin{itemize}

\item $\alpha_j$-modes: $\omega=-(j+2)$ and $0\leq j\leq N-2$, where each $j$-mode has a degeneracy of $2j+1$. Their masses are given by
\beq%
M^2_{\alpha_j}= \frac12\lambda_{eff}\phi^2 (j+2)(j+3)-2\kappa_{eff}\phi (j+2)+m^2\,.
\eeq%
 The single
$\alpha_{j=0}$ mode corresponds to the adiabatic mode itself. There is therefore, $(N-1)^2-1$ of spectators from the scalar $\alpha$-sector.

\item $\beta_j$-modes: $\omega=j-1$ and $1\leq j\leq N$. Degeneracy of each $\beta_j$-mode is $2j+1$ and hence there are $(N+1)^2-1$ of $\beta$-modes. Mass of $\beta_j$ mode is%
\beq
M^2_{\beta_j}=\frac12 \lambda_{eff}\phi^2 (j-1)(j-2)+2\kappa_{eff}\phi (j-1)+m^2\,.
\eeq%

\end{itemize}

\paragraph{\small{Spectrum of gauge field spectators}} The spectrum of gauge fields can be also derived expanding the action \eqref{action} to second order in $A_\mu$, keeping the $\Phi_i$ in the $SU(2)$ sector. For that one should note that  the eigenvalue problem $[J_i,[J_i, X]]=\omega X$, which has eigenvalues
$j(j+1)$, and thus the gauge field spectators mass spectrum could be expressed as
\beq%
M^2_{A,j}=\frac{\lambda_{eff}}{4}\phi^2 j(j+1)\,.
\eeq%
 $j=0$ mode remains massless and corresponds to the $U(1)$ sector in the $U(N)$ matrices. The fact that it remains massless illustrates the fact that we have freedom in choosing the center of mass of the system. The degeneracy of the vector field modes is hence $3(2j+1)$ for $j\geq 1$ modes and is two for $j=0$ mode. We hence have effectively $3N^2-1$ spin one vector field modes, all except for two are massive. These gauge modes substitute $N^2-1$ ``zero modes'', which are unphysical  in the gauged theory. The zero modes were massless in the minimum  of the potential which justifies the appellation.

Overall we have $(N-1)^2-1$  $\alpha$-modes, $N^2+2N$ $\beta$-modes and $3N^2-1$ vector field modes that altogether form $5N^2-1$ dof of the model besides the $SU(2)$  direction.

\section{Shortcomings of $\MM$-flation}

With Matrix Inflation coupled minimally  to gravity, one can show that in the region $\phi>\mu$, the predictions of inflation have already been ruled out by PLANCK 2018 data \cite{Ashoorioon:2014jja, Akrami:2018odb}. In the limit that $\mu/M_{P}\rightarrow 0$, the predictions of the model in the $n_{_{S}}-r$ plane approaches the predictions of $\lambda\phi^4$ theory, which was already ruled out by WMAP 2005 data. As $\mu/M_{P}$ becomes larger, the tensor-to-scalar $r$ gets smaller and in the limit of $\mu/M_{P}\rightarrow \infty$,  the predictions of the model approaches that of $m^2\phi^2$, which was in tension with the PLANCK 2013 data. For the central value of scalar spectral index from PLANCK2018 data, \cite{Akrami:2018odb}, $n_{{}_S}=0.9649$, $\mu\approx 95.65~M_{P}$ and the predicted tensor-to-scalar ratio, $r$ is $0.1581$. When the BICEP2 result was announced, the signal from this region of the potential was in the sweet spot of the $n_{_{S}}-r$ plane, $r\sim 0.2$ (with the PLANCK 2013 value for the central value of $n_{_{S}}$). However after the dust settled, we now know that this region of parameter space is ruled out, unless one tries to suppress the tensor-to-scalar ratio by modifying the quantum fluctuations of the perturbations as in \cite{Ashoorioon:2013eia}. In this region assuming that the bare quartic coupling to be of order one, {\it i.e.} $\lambda\equiv 8\pi g_{{}_S}\simeq 1$ \footnote{The Yang-Mills perturbative coupling $\alpha_{{}_{\rm YM}}$ is then related to $\lambda$ as $\alpha_{{}_{\rm YM}}\equiv \frac{g_{{}_{\rm YM}}^2}{4\pi}=\frac{\lambda}{8\pi}= g_{{}_{S}}\ll 1$, so that perturbative expansion is reliable.}, one needs about $109850$ D3 branes. The regions (b) and (c), on the other hand, are concave potentials and are better suited to match with the limits the latest PLANCK data put on the tensor-to-scalar ratio. For the central value of the spectral index from \cite{Akrami:2018odb}, $n_{{}_S}=0.9649$, the predicted tensor-to-scalar ratio, $r$ is $0.055$, which is still within the $2\sigma$ region in the $n_{{}_S}-r$ plane in the latest PLANCK 2018 results. With bare coupling of order one in this region, the required number of D3 branes is reduced to $N=54820$ D3 branes, which is expected to be too many to be implemented in a realistic string theory realization, when compactified \footnote{In order to do compactifications, one probably needs to turn on the $H_3$ flux, which couples the flux $\tilde{F_5}$ emanated from the stack of D3 branes to the strength of the RR 6-form flux (which could be generated by a distribution of D5 branes) through the Bianchi identity $ d \tilde{F}_5=H_3\wedge F_3 $.}.

Another problem that is related to the large number of D3 branes in the minimal setup is the running of SU(N) non-Abelian gauge theory, when one runs from the string scale down to the scale of inflation. Of course the scale of inflation is very close to the GUT scale, which should only be one or two orders of magnitude below the string scale. Still due to the large number of D3-branes this could be problematic. To see that, one should note that for a $SU(N)$ gauge theory with three real scalars, the running of perturbative Yang-Mills coupling\footnote{When two fermions are included, the factor $-\frac{19 N}{12\pi}$ will change to $-\frac{11 N}{12\pi}$.}, $\alpha_{{}_{\rm YM}}\equiv \frac{g_{{}_{\rm YM}}}{4\pi}$,
\beq
\beta_{{}_{\rm YM}}\equiv \frac{d\alpha_{{}_{\rm YM}}}{d\ln \mu}=-\frac{19 N}{12\pi}\alpha_{{}_{\rm YM}}^2\,,
\eeq
which determines the perturbative Yang-Mills coupling as a function of energy scale $\mu$,
\beq
\alpha_{{}_{\rm YM}}(\mu)=\frac{\alpha(\mu_0)}{1+\frac{19 N}{12\pi}\ln\left(\frac{\mu}{\mu_0}\right)}\,.
\eeq
Taking the separation between $\mu_0$, the string length, and $\mu$, the inflationary scale, to be one order of magnitude, this suggests that
\beq
\alpha(\mu_0)=g_{{}_S}\lesssim \frac{1}{1+\frac{19 N}{12}\ln 10}\,,
\eeq
and for $N\sim 5\times10^4-10^5$, this suggests that
\beq
g_{{}_S}\lesssim 10^{-6}\,,
\eeq
which although lessens the fine-tuning required to lower the inflaton's self-coupling to one in $10^6$. The rest of the suppression in inflaton's coupling can be provided with about $10^3$ D3 branes.

The $SU(2)$ configuration is not a local attractor in the whole range of $\phi>0$ for all spectator directions. Although gauge spectators, except for the $j=0$ mode\footnote{The masslessness of $j=0$ is the representative of our freedom in choosing the center of mass of the system.}, all have positive mass squared. However the $\alpha$ and $\beta$ modes masses can change sign. Indeed in the range $\phi_2<\phi<\phi_1$,
\beqa
\phi_{1}= \frac{-3\omega+\sqrt{5\omega^2+4\omega}}{2(\omega^2-\omega)}\mu\,,\nonumber\\
\phi_{2}= \frac{-3\omega-\sqrt{5\omega^2+4\omega}}{2(\omega^2-\omega)}\mu\,,
\eeqa
where for $\alpha$-modes, $\omega=-(j+2)$ where $1\leq j\leq N-2$, and for $\beta$-modes where $\omega=j-1$ and $1\leq j\leq N$, the scalar spectator modes can become tachyonic. In region (a), $\phi>\frac{\mu}{2}$, all of these modes are curved up and therefore the $SU(2)$ configuration is the local attractor in this region of the potential. In region (b) the mode $\omega=-3$, which corresponds to $j=1$, $\alpha$-mode,  becomes tachyonic for values of $\phi\lesssim 0.6144\mu$. For value of $\mu\simeq 41.87 M_{P}$, which yields the mean value of scalar spectral index of the PLANCK 2018 data, this happens before the CMB scales exit the horizon. In particular for such a value of $\mu$, this happens before $\phi_{60}$. The number of e-folds one would get with such value of $\mu$ is about 109 e-folds. Although this is more than what is needed to solve the problems of the standard Big Bang cosmology, eternal inflation \cite{Linde:1986fc} is no longer possible in this region of parameter space. In region (c), the spectator modes with $-79\leq\omega \leq -3$ all become tachyonic at some point during inflation and so the trajectory cannot lead to stable inflationary trajectory during inflation. Contrary to the previous understandings, the mass squared of some of these unstable spectators becomes smaller than $-H^2$ and can terminate inflation along the $SU(2)$ direction abruptly.

The mass of the spectator fields are dependent on the value of the inflaton. Indeed they vary as the inflaton rolls. In region (a) of the potential {\it most} of them start from larger values and then towards the end of inflation, when $\phi$ approaches $\mu$, they become lighter \footnote{Of course a smaller fraction of them becomes  heavier during the inflaton's excursion.}. This is another representation of the distance conjecture of Vafa and Ooguri \cite{Ooguri:2006in, Obied:2018sgi, Ooguri:2018wrx,Garg:2018reu} that by displacement in the moduli space, a tower of scalar fields become light. As mentioned previously, the amount of excursion of the physical inflaton, assuming bare coupling of order one, is $\sim 10^{-6}~M_{\rm P}$. One would not see exponential lightening of the moduli fields, but rather a polynomial behavior. In region (b), one instead notices that  these moduli will become heavier as inflation progresses. In region (c), the situation is similar to region (a)
and in fact some of the modes become tahyonic, as explained above.

The fact that spectator fields' masses are dependent on the vev of inflaton, allows for $\MM$-flation to have a ``potential'' embedded preheating mechanism. During inflation, the variation of these masses are, however, small and slow-roll suppressed. In \cite{Ashoorioon:2009wa}, it was shown that the amount of particles produced during inflation, despite the large degeneracy of high $j$ modes for $\alpha$ and $\beta$ modes is small. Although the focus was on the ungauged $\lambda\phi^4$ inflationary model in the argument that was presented there  \cite{Ashoorioon:2009wa}, the same argument and computations could be applied to the case of gauged $\MM$-flation. For region (a) and (b), the argument in fact gets further amplification due to the massiveness of the spectator modes around $\phi=\mu$ vacuum. Large $j$ modes will be too heavy to be created adiabatically and the degeneracy of light modes is too small to have a substantial backreaction on inflationary background. In region (a), although some of the modes become tachyonic during inflation, since this happens only when the mode goes outside the horizon, the argument of \cite{Ashoorioon:2009wa} will still remain applicable and valid.

Although the energy density of the produced particles is too small to derail inflation, they can become a dominant effect after the slow-roll condition is violated to drain the energy of the inflaton, specially since the masses of the spectators are dependent on inflaton and they can be produced non-adiabatically. This can act as an effective way of depleting the energy of inflaton and transferring it to the standard model sector. However, as it was shown in \cite{Ashoorioon:2013oha}, the large $j$ $\alpha$ and $\beta$ modes will be again too massive to be produced non-adiabatically. For small $j$ modes on the other hand, although the adiabaticity condition is violated, their masses are such that they fall out of the instability bands quickly and are not effective in transferring the energy of the inflaton.  Of course, it is not necessary that the reheating comes along with the inflationary sector, but it could have been quite interesting that this feature of Matrix inflation could be put to use. In region (c) of the potential, however, these modes can have zero (gauge modes) or light (scalar modes) bare value of mass and could easily be produced non-adiabatically during the oscillations of the inflaton \footnote{In fact, the scalar modes not only become massless but also tachyonic for a small region around $\phi=0$ and can contribute copiously to the particle production and reheating of the universe.}. One way preheating could have worked out, is that the inflaton passes over the barrier from  $\phi=\mu$ to $\phi=0$ and start oscillating around $\phi=0$. However the two vacua in $\MM$-flation are too far apart, as $\mu$ has to take super-Planckian values to match with the CMB limits on the spectral index.

As we will see, assuming that inflaton matrices are non-minimally coupled to gravity, one can mitigate all of these problems.

\section{The Setup of Non-Minimal M-flation}
\label{section:setup}

\subsection{Non-Minimal Coupling: Motivation}

Although the aim of this article is to investigate  the effect of non-minimal coupling to gravity for the $\MM$-flationary predictions, one may wonder if such non-minimal coupling to gravity can be motivated from a more fundamental top-down approach.

We have assumed so far that once we compactify and come down from 10 dimension to 4 dimension, the gravity remains decoupled from the matter sector. However this is not necessarily true. In presence of matter, it is known that the gravitational action can get renormalized. In fact in presence of a scalar field, it was shown in \cite{Ashoorioon:2011aa} that the loop correction to the graviton-scalar-scalar vertex generates a term proportional to $\xi \frac{\Lambda^2}{M_P^2} R\phi^2$, where $\xi<0$ and $|\xi|\lesssim \mathcal{O}(1)$. If the cutoff is taken to be $M_{P}$, this would in general create a mass of order $H^2$ for the inflaton, which would then result in the $\eta$ problem. This is what was called quantum $\eta$-problem in \cite{Ashoorioon:2011aa} . In many-field models of inflation, where a lot of scalar species are involved, like N-flation \cite{Dimopoulos:2005ac} or original ungauged M-flation \cite{Ashoorioon:2009wa}, since there are more than one light fields involved, they contribute to the renormalization of the Planck mass and lower the cutoff of the theory to $\Lambda$, where
\beq
\Lambda=\frac{M_P}{\sqrt{N_l}}\,,
\eeq
in which $N_l$ is the number of light species with $m<\Lambda$. The previous correction to the three vertex operator, would then acquire a suppression of $1/N_l$ in the induced non-minimal coupling factor. As we will see in section \ref{UVcutoff}, in gauged (non-)$\MM-$flation though, the presence of vector modes, can keep the cutoff around the Planck scale, $M_P$. Therefore one may expect to obtain $\xi\sim -1$ in such models.

Another way one can see that the generation of sizable non-minimal to gravity is inevitable, is by looking at  how the  moduli stabilization procedure in the string theory realization of inflation works out. For example in the $\mathbb{K}$L$\mathbb{M}$T \cite{Kachru:2003sx}, assuming that the superpotential is dependent only on the $\rho$ modulus, the stabilization of volume modulus leads to terms proportional to conformal coupling to gravity, $\xi R\phi^2$ in  4 dimensions, with $\xi=-\frac{1}{6}$, which in turn causes the famous $\eta$-problem. One way to avoid this problem is by assuming that the superpotential is dependent on the position of the D3 brane in the throat \cite{McAllister:2005mq}. Then the contribution from such a dependence was tuned to make the mass term for the inflaton small. However, in principle, naturally this coefficient is much larger than one and  large non-minimal coupling of inflaton to gravity is produced. Assuming that the superpotential depends on the position moduli of the D3 brane in  $\mathbb{K}$L$\mathbb{M}$T setup,\beq
W(\rho,\phi)=W_0+g(\rho) f(\phi)\,,
\eeq
where $\rho$ is proportional to the volume modulus and $\phi$ is the D3 brane position moduli. $g(\rho)$ is an arbitrary function and $f(\phi)=1+\delta~\phi^2$, it turns out that such a mass term is proportional to
\beq
m_{\phi}^2= 2H^2 \left( 1+\frac{V_{\rm AdS}}{V_{\rm dS}}(2\beta^2-\beta)\right)\,,
\eeq
where
$\beta=\delta \frac{g(\rho)}{g'(\rho)}$. Now if $\frac{V_{\rm AdS}}{V_{\rm dS}}\gg 1$, as it happens in the case analyzed in \cite{Kachru:2003aw}, and $\beta\neq 0$ or $1/2$, the resulted mass term will be large. In particular, if $0<\beta<\frac{1}{2}$ the resulted $m_{\phi}^2$ would be negative and thus the non-minimal coupling, $\xi$, could become large. In addition, it is natural to expect that in presence of a stack of D3-branes, we can have cumulative effects that enhances such non-minimal couplings to gravity and justify $\xi R\phi^2$, with $\xi\gg 1$. As we will see such non-minimal coupling to the gravity in the case of Matrix inflation can amend the problems discussed previously to be associated with $\MM$-flation.

\subsection{Non-Minimal M-flation}

Following the above argument, we {\it posit} that the effect of dependence of the superpotential on the position moduli of the D3 branes, modifies the action for Matrix inflation in the following manner
\begin{align}
 S_{{}_{\rm Non-\mathbb{M}-flation}}= & \int d^{4}x\sqrt{-g}\bigg[\frac{1}{2}\left(1+\xi\sum\limits _{i=1}^{3}\mathrm{Tr}\left(\Phi_{i}^{2}\right)\right)R-\frac{1}{2}\sum\limits _{i=1}^{3}{\rm Tr}\left(D_{\mu}\Phi_{i}D^{\mu}\Phi_{i}\right)
 \nonumber
 \\
 &-V\left(\Phi_{i},\left[\Phi_{i},\Phi_{j}\right]\right) -\frac{1}{4}{\rm Tr}\left(F_{\mu\nu}F^{\mu\nu}\right)\bigg]\,.
 \label{S-NonMflation}
\end{align}
To simplify the background dynamics, again we use the same trick as in $\MM$-flation and we go to the $SU(2)$ sector, which leads to a scalar field with non-canonical kinetic term. The process of canonicalization of the kinetic term for the field, provides us with a potential with suppressed cubic and quartic couplings. The scalar field is non-minimally coupled to gravity. The action takes the following form
\begin{equation}
 \label{SJphi}
 S_{{}_{\rm Non-\mathbb{M}-flation}}^{\rm J}=\int d^{4}x\sqrt{-g}\left[\frac{1}{2}\left(1+\xi\phi^{2}\right)R+\frac{1}{2}\left(\frac{d\phi}{dt}\right)^{2}-V_{0}(\phi)\right],
\end{equation}
where the potential is defined as in \eqref{Vphi}. The effective couplings get dressing factors as in eqs. \eqref{lameff}.

One can  we make the following conformal transformation to move from the Jordan frame to the Einstein frame, where the gravity looks like the Einstein-Hilbert term
\begin{equation}
 \label{gtildemunu}
 \tilde{g}_{\mu\nu}=\Omega^{2}g_{\mu\nu}\,,
\end{equation}
where tilde is applied to identify the metric in the Einstein frame. The conformal factor of this transformation is
\begin{equation}
 \label{Omega}
 \Omega^{2}=1+\xi\phi^{2}\, .
\end{equation}
Due to the conformal transformation, the kinetic term of the inflaton field becomes non-canonical again. It is convenient to define the new scalar field $\chi$ which is related to $\phi$ by
\begin{equation}
 \label{dchidphi}
 \frac{d\chi}{d\phi}=\sqrt{\frac{\Omega^{2}+6\xi^{2}\phi^{2}}{\Omega^{4}}}\,.
\end{equation}
At the end of the day, the action in the Einstein frame turns into
\begin{equation}
 \label{SE}
 S^{\rm E}_{{}_{\rm Non-\mathbb{M}-flation}}=\int d^{4}x\sqrt{-\tilde{g}}\left[\frac{1}{2}\tilde{R}+\frac{1}{2}\left(\frac{d\chi}{d t}\right)^{2}-U(\chi)\right]\,,
\end{equation}
where now the potential in the Einstein frame takes the form
\begin{equation}
 \label{UchiV0phi}
 U(\chi)=\frac{V_{0}\left(\phi(\chi)\right)}{\Omega^{4}\left(\phi(\chi)\right)}\,.
\end{equation}
One can explicitly work out the relation between $\chi$ and $\phi$ using eq. (\ref{dchidphi}), which is
\begin{align}
 \chi=f(\phi)= & \sqrt{\frac{6\xi+1}{\xi}}\ln\left[\sqrt{\xi\left(6\xi+1\right)\left(\xi(6\xi+1)\phi^{2}+1\right)}+\xi\left(6\xi+1\right)\phi\right]
 \nonumber
 \\
 & -\sqrt{6}\tanh^{-1}\left[\frac{\sqrt{6}\xi\phi}{\sqrt{\xi(6\xi+1)\phi^{2}+1}}\right]\, .
 \label{chiphi}
\end{align}
The inverse function $f^{-1}$ can be used to express $\phi$ in terms of $\chi$,
\begin{equation}
 \label{phichiexact}
 \phi=f^{-1}(\chi)\,.
\end{equation}
The function $f^{-1}$, although is an implicit function, can be used to determine $\chi$ exactly in terms of $\phi$ without resorting to any approximation. As we will see in appendix \ref{section:Higgs}, in the case of Higgs inflation, we will obtain a tiny but measurable correction to the predictions of the model and therefore throughout this paper, in the treatment of $\MM$-flation, we will only use the implicit but exact form of $\chi$ and the potential in the Einstein frame. Using this and eq. (\ref{Omega}) in eq. (\ref{UchiV0phi}), the potential for the $\chi$ field in the Einstein frame becomes
\begin{equation}
 \label{Uchi}
 U(\chi)=\frac{\lambda_{\mathrm{eff}}\left(f^{-1}(\chi)\right)^{2}\left(f^{-1}(\chi)-\mu\right)^{2}}{4\left(1+\xi\left(f^{-1}(\chi)\right)^{2}\right)^{2}}\,.
\end{equation}
 \begin{figure}[t]
  \centering
\includegraphics[scale=0.7]{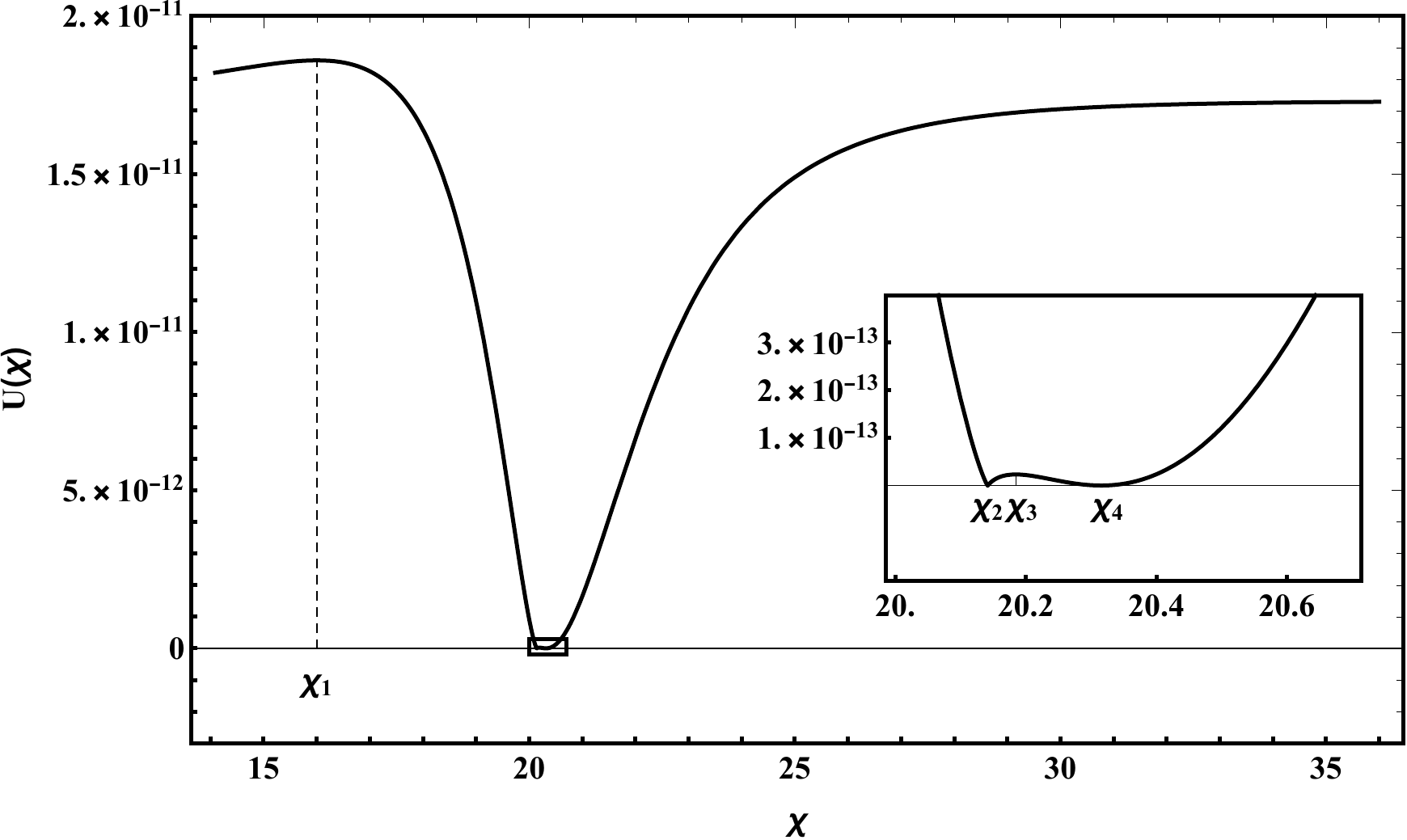}
\caption{The potential for the canonical Einstein frame inflaton, $\chi$. The lopsided form of the potential becomes enhanced and is no longer symmetric around $\chi=\chi_3$ (which corresponds to $\phi=\mu/2$ in the Jordan frame potential). The region $\chi>\chi_4$, corresponding to region $\phi>\mu$, flattened further and suitable for PLANCK 2018 compatible inflation. One can also inflate in the region $\chi_2<\chi<\chi_3$, but not for all values of $\xi$ and $\mu$.  }\label{figure:Uchi2}
\end{figure}
The potential is shown in fig. \ref{figure:Uchi2}. As we see in the figure, this potential has two minima at
\begin{align}
 \label{chi2}
 \chi_{2}=f(0)= & \sqrt{\frac{1}{\xi}+6}\ln\sqrt{\xi(6\xi+1)},
 \\
 \chi_{4}=f(\mu)= & \sqrt{\frac{6\xi+1}{\xi}}\ln\left[\sqrt{\xi\left(6\xi+1\right)\left(\xi\mu^{2}(6\xi+1)+1\right)}+\xi\mu\left(6\xi+1\right)\right]
 \nonumber
 \\
 & -\sqrt{6}\tanh^{-1}\left[\frac{\sqrt{6}\xi\mu}{\sqrt{\xi\mu^{2}(6\xi+1)+1}}\right].
 \label{chi4}
\end{align}
The potential (\ref{Uchi}) also has two local maxima at $\chi_1$ and $\chi_3$ that can be determined too, by setting $U'(\chi)=0$.

Now we proceed to studying the  non-$\MM$-flation with the potential given in Eq. (\ref{Uchi}). Let us introduce the first and the second slow-roll parameters,
\begin{align}
 \label{epsilon}
 \epsilon & =\frac{M_P^2}{2}\left(\frac{U'(\chi)}{U(\chi)}\right)^{2}\,,
 \\
 \label{eta}
 \eta & =M_P^2 \frac{U''(\chi)}{U(\chi)}\,.
\end{align}
These parameters are much less than unity during the slow-roll inflation, and the end of inflation is marked by $\epsilon = 1$. In the study of inflation, it is conventional to quantify the inflaton dynamics in terms of the e-folding defined as $N_e \equiv\ln\left(a_\mathrm{end}/a(t)\right)$, where $a(t)$ is the scale factor that its value at the end of inflation is shown by $a_e$. The largest scale in the CMB corresponds to the scale that exit the horizon about $N_e=50$ to $N_e=60$ depending on the scales of inflation and reheating temperature. In the standard inflationary scenario and in the slow-roll approximation, the scalar field evolution could be related to the number of e-folds through the equation \begin{equation}
 \label{dchidNe}
 \frac{d\chi}{dN_e}\approx M_P^2 \frac{U'(\chi)}{U(\chi)}\,.
\end{equation}
In the framework of the Einstein gravity and with a canonical scalar field as the inflaton, the scalar and tensor power spectra are given respectively by
\begin{align}
 \label{Ps}
 \mathcal{P}_{\rm {}_S} & =\frac{U(\chi)}{24\pi^{2}M_P^4\epsilon},
 \\
 \label{Pt}
 \mathcal{P}_{\rm {}_T} & =\frac{2U(\chi)}{3 M_P^4\pi^{2}}.
\end{align}
The reported value for the amplitude of the scalar perturbations at the horizon crossing according to the Planck 2018 TT,TE,EE+lowE+lensing data is $\ln\left(10^{10}\mathcal{\mathcal{P}}_{s}\right)=3.044\pm0.014$ (68\% CL) \cite{Akrami:2018odb}, and we use this constraint to determine the parameter $\lambda_{\mathrm{eff}}$ in our analysis. In the setting of the standard inflation, the scalar spectral index and the tensor-to-scalar ratio can be worked out in terms of the slow-roll parameters,
\begin{align}
 \label{ns}
 n_{{}_S} & =1-6\epsilon+2\eta\,,
 \\
 \label{r}
 r & =16\epsilon\,.
\end{align}
These quantities are observables which are used to discriminate between inflationary models. We estimate these observables in our non-$\mathbb{M}$-flation model and compare the results with the results of the PLANCK 2018 data \cite{Akrami:2018odb}.

The reader may wonder why we have tried to compute the scalar curvature perturbations in the Einstein frame and not the Jordan one, where the original action is defined. It should be noted that in the model at hand, in both frames, inflation is essentially driven by a single scalar field, the one which corresponds to the radius of the single polarized giant D5 branes. As we will see in section \ref{section:isocurvature}, the other spectator fields are frozen classically in both frames and only contribute quantum mechanically in the form of isocurvature perturbations. The metric of the field space in the Jordan frame is  $\delta_{IJ}$ and the trajectories in the field spaces in both frames, are straight lines. In such a case, the equivalence between the Jordan and Einstein frames quantities for the curvature perturbations has been established, please see \cite{White:2012ya, Postma:2014vaa}. We have studied the cosmological perturbations in the Einstein frame because tracking the computations was easier in this frame. To which metric the ordinary matter is coupled depends on how one embeds the current inflationary sector in a full-fledged string theory framework. In this paper we assumed that the normal matter field is coupled to the Einstein frame metric and thus curvature perturbations in the Einstein frame is conserved. This is also supported by the fact that after inflation, the inflaton settles in one of its minima and thus there will be no further evolution of the Planck mass after inflation.

\subsubsection{Region $\chi>\chi_4$}

\begin{figure}[t]
  \centering
\includegraphics[scale=0.3]{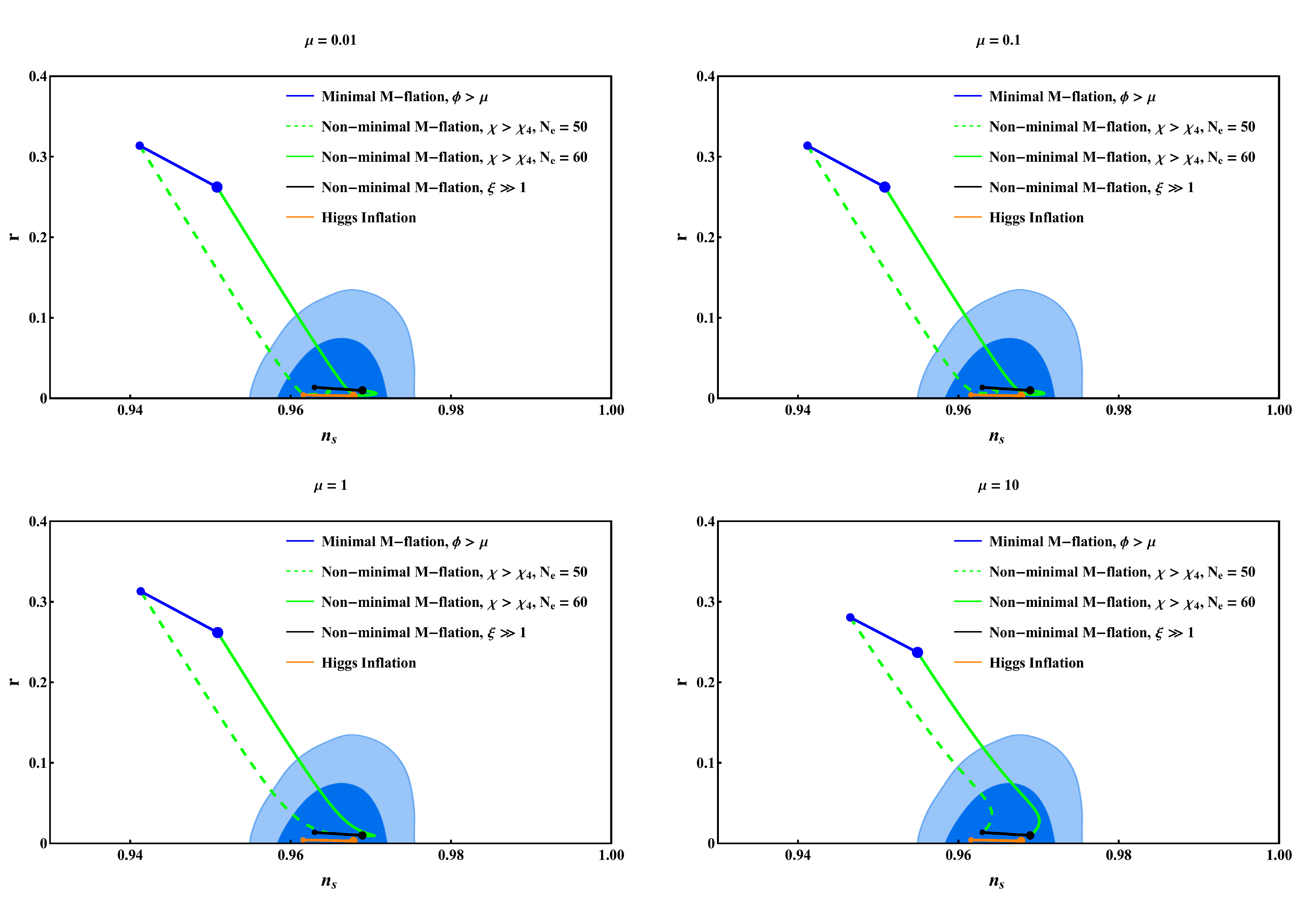}
\caption{The predictions of non-$\MM$-flation in region $\chi>\chi_4$ for various values of $\mu$, symmetry-breaking vev. As $\xi$ increases from zero, the predictions of the model rolls inside the $2 \sigma$ region of PLANCK 2018 data.}
\label{figure:rnsmu-chi-bigger-chi4}
\end{figure}

First, we focus on the region $\chi > \chi_4$ of the potential (\ref{Uchi}) which now has obtained a plateau shape in the Einstein frame. This region corresponds to the part $\phi > \mu$ of the Jordan frame potential \eqref{Vphi}.
The predictions of our model in this region and with some typical values $\mu$ have been presented in fig. \ref{figure:rnsmu-chi-bigger-chi4}. We have also specified the $68\%$ and $95\%$ C. L. marginalized joint regions of PLANCK 2018 TT,TE,EE+lowE+lensing data \cite{Akrami:2018odb} in the figures. From these figures, we conclude that the predictions of the non-$\mathbb{M}$-flation model in this region can become compatible with the PLANCK 2018 TT,TE,EE+lowE+lensing data \cite{Akrami:2018odb}. In the figures, the predictions of our models are shown in green. The predictions of $\mathbb{M}$-flation models are shown by blue circles. The plots for the predictions of the non-$\mathbb{M}$-faltion models were made by varying the non-minimal parameter. The dashed and solid curves, respectively, correspond to the CMB scales exiting the horizon at $N_e=50$ or  $60$ e-folds before the end of inflation. The prediction of non-$\mathbb{M}$-flation model in the limit of $\xi \gg 1$ is shown by black circles. As we show in appendix \ref{section:xigg1}, the predictions of all non-$\MM$-flationary models approach a unique point in the $n_{{}_{S}}-r$ plane, in this limit. We emphasize that the predictions of the models were extracted using the exact implicit form of the function $f^{-1}(\chi)$. We did not use the large $\xi$ approximation used in Higgs inflation \cite{Bezrukov:2007ep}, because we also look at the region of $\xi\lesssim 1$.   As $\xi$ increases, the predicted tensor-to-scalar ratio $r$ decreases and $n_{{}_S}$ enhances. For $\mu<M_{P}$, there are two turning points, in the first, $r$ bounces up and in the second one $n_{{}_S}$ starts to decrease. For $\mu \gtrsim M_{P}$, it seems that only the latter exists. We see that the results of the non-$\mathbb{M}$-flation models can become consistent with the PLANCK 2018 data for a range of values of $\xi$. In the figures, we also have demonstrated the prediction of the Higgs inflation model \cite{Bezrukov:2007ep}, using the exact implicit function $f^{-1}(\chi)$. Even in the limit of $\xi\gg 1$, the predictions of Higgs inflation is slightly different from what one would obtain using such an approximation, please see the appendix \ref{section:Higgs}. This is due to the fact that the potential of $\mathbb{M}$-flation in the Einstein frame, contrary to the Higgs potential, is asymmetric. In the Higgs potential, the value of $\mu$ is fixed to $246$ GeV and the quartic coupling to $\lambda\simeq 0.129$. However in the case of non-$\mathbb{M}$-flation, even after fixing $\mu$, depending on the value of $\xi$, the required value of $\lambda$ changes and therefore the predictions of the model can still vary in the $n_{{}_S}-r$ plane, as shown in fig. \ref{figure:rnsmu-chi-bigger-chi4}.
\begin{table}[t]
  \centering

  \scalebox{1.0}{
  \begin{tabular}{lccccc}
    \hline
    \hline
    $\mu\qquad$ & $\qquad\xi\qquad$ & $\qquad\qquad\lambda_{{\rm eff}}\qquad\qquad$ & $\qquad n_{s}\qquad$ & $\qquad r\qquad$ & $\qquad N\qquad$\tabularnewline
    \hline
    $0.01$ & $500$ & $1.599\times10^{-4}$ & $0.9704$ & $0.0049$ & $37$\tabularnewline
    $0.01$ & $1520$ & $1.844\times10^{-3}$ & $0.9707$ & $0.0061$ & $16$\tabularnewline
    $0.1$ & $1000$ & $1.295\times10^{-3}$ & $0.9691$ & $0.0096$ & $18$\tabularnewline
    $1$ & $100$ & $1.325\times10^{-5}$ & $0.9690$ & $0.0098$ & $84$\tabularnewline
    $10$ & $572.2$ & $4.345\times10^{-4}$ & $0.9690$ & $0.0098$ & $26$\tabularnewline
    $100$ & $284.8$ & $1.0766\times10^{-4}$ & $0.9690$ & $0.0098$ & $42$\tabularnewline
    \hline
  \end{tabular}
  }
  \caption{The inflationary observables including the scalar spectral index $n_{_{S}}$ and tensor-to-scalar ratio $r$ in the non-$\mathbb{M}$-flation model for some typical values of the parameters $\mu$ and $\xi$. Also, the derived values of the parameter $\lambda_{{\rm eff}}$ and number of the D3-branes $N$ are presented in the table. The quantities in this table are evaluated with the horizon exit e-fold number $N_e = 60$.}
  \label{table:nsrN}
\end{table}

\begin{table}[b]
  \centering
  \scalebox{1.0}{
\begin{tabular}{lccc}
\hline
\hline
$\mu$ & $\qquad N_{e} \qquad$ & $\qquad$ 68\% CL $\qquad$ & $\qquad$ 95\% CL $\qquad$\tabularnewline
\hline
\multirow{2}{*}{$0.01$} & \multicolumn{1}{c}{$50$} & $\xi\gtrsim3.0\times10^{-2}$ & $\xi\gtrsim1.1\times10^{-2}$\tabularnewline
 & $60$ & $\xi\gtrsim6.2\times10^{-3}$ & $\xi\gtrsim3.2\times10^{-3}$\tabularnewline
\hline
\multirow{2}{*}{$0.1$} & $50$ & $\xi\gtrsim2.9\times10^{-2}$ & $\xi\gtrsim1.1\times10^{-2}$\tabularnewline
 & $60$ & $\xi\gtrsim6.3\times10^{-3}$ & $\xi\gtrsim3.2\times10^{-3}$\tabularnewline
\hline
\multirow{2}{*}{$1$} & $50$ & $\xi\gtrsim2.5\times10^{-2}$ & $\xi\gtrsim1.1\times10^{-2}$\tabularnewline
 & $60$ & $\xi\gtrsim6.2\times10^{-3}$ & $\xi\gtrsim3.1\times10^{-3}$\tabularnewline
\hline
\multirow{2}{*}{$10$} & $50$ & $\xi\gtrsim1.3\times10^{-2}$ & $\xi\gtrsim5.1\times10^{-3}$\tabularnewline
 & $60$ & $\xi\gtrsim5.0\times10^{-3}$ & $\xi\gtrsim1.8\times10^{-3}$\tabularnewline
\hline
\end{tabular}
  }
   \caption{Range of the coupling parameter $\xi$ for which the non-$\mathbb{M}$-flation model is compatible with the 68\% or 95\% CL constraints of the Planck 2018 data \cite{Akrami:2018odb} in the $r-n_s$ plane.}
  \label{table:xi}
\end{table}

It is useful to estimate the exact number of D3-branes needed for various values of $n_{_{S}}$ and $r$ in the $2\sigma$ region of PLANCK 2018 data, for some typical values of $\mu$ and $\xi$. The results are tabulated in Table \ref{table:nsrN}. In the last column of this table, we present the number of D3-brane needed in our non-$\mathbb{M}$-flation model in order to match the data, assuming that the undressed quartic coupling $\lambda\sim \mathcal{O}(1)$. We see in table \ref{table:nsrN} that with $\xi \sim {\rm few}\times 100$, the number of D3-branes are $N \lesssim 100$. Therefore, one can conclude that we can reduce the number of D3-branes in the setup of non-$\mathbb{M}$-flation in comparison with the $\mathbb{M}$-flation model that required $N \sim 10^4-10^5$, by a factor of $10^2-10^3$. This will be important once one attempts to compactify the model, since large number of D3-branes, in presence of fluxes needed for moduli stabilization and compactification, can backreact on the background pp-wave geometry.  One should note that not necessarily large values of $\xi$ are needed to reconcile the predictions of non-$\MM$-flation with the PLANCK 2018 data. As it has been tabulated in table \ref{table:xi}, values of $\xi$ from $10^{-3}$ to $10^{-2}$ can serve this purpose. As we will see later, with such values of non-minimal coupling one would get a partial reheating and transfer of the energy of the $SU(2)$ direction inflaton to spectator sector. However with such a small value of $\xi$, the required number of D3 branes remains of the same order as before, {\it i.e.} $10^{4}-10^{5}$.
\begin{figure}
 \centering
 \includegraphics[scale=0.4]{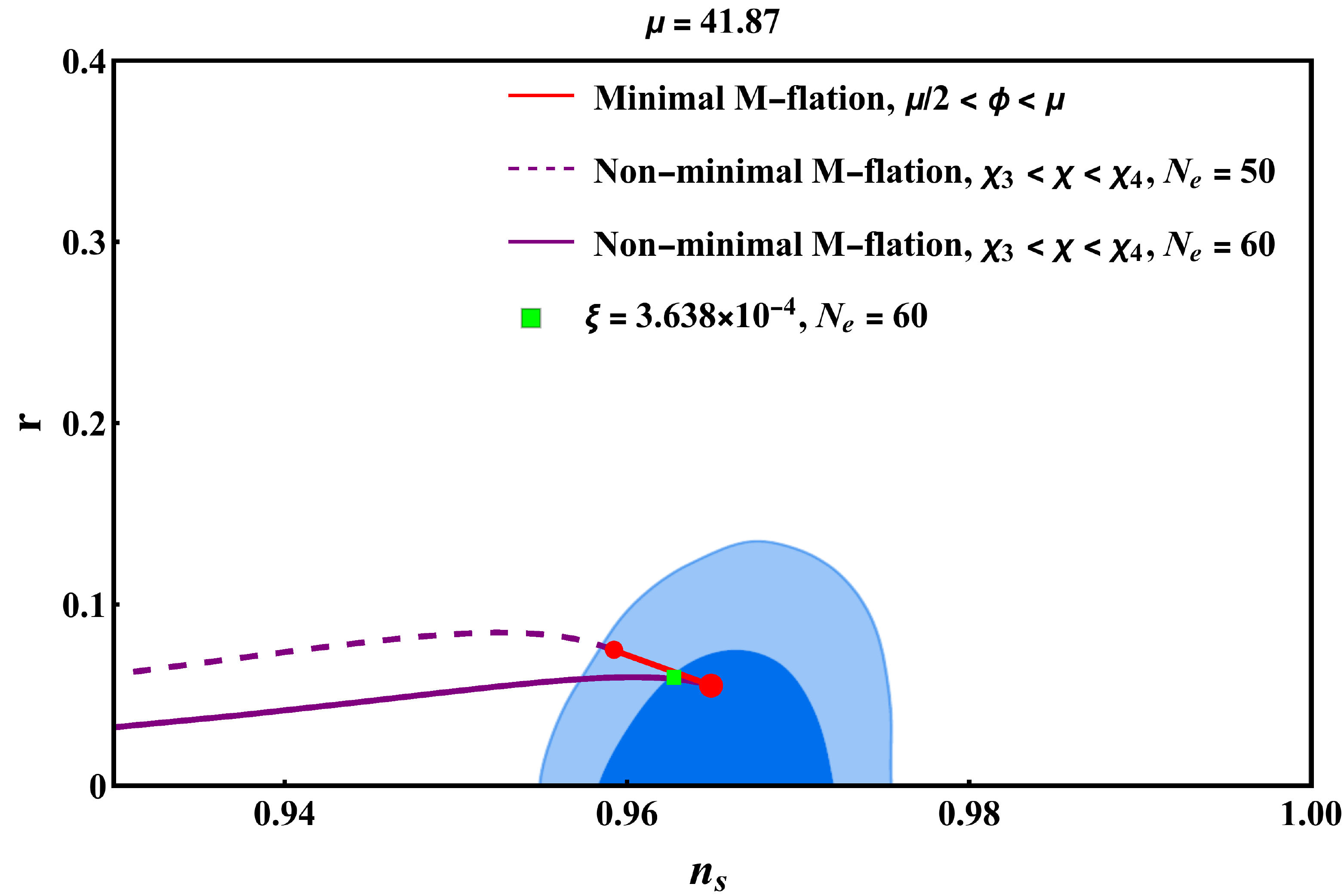}
 \caption{This figure depicts the evolution of non-$\MM$-flation predictions in the hilltop region $\chi_2<\chi<\chi_4$. Enhancing the value of $\xi$ slightly one runs out of the PLANCK $2\sigma$ viable region in the $n_{{}_S}-r$ plane. Also the length of inflation shortens. The green square dot corresponds to the value of $\xi$ for which only the stable part of the potential in this region supports 60 e-foldings of inflation.}
 \label{figure:rnsmu41p87chi4}
\end{figure}

\subsubsection{Region $\chi_2<\chi<\chi_4$}

Slow-roll inflation can also take place in the region $\chi_2 < \chi < \chi_4$ if the parameter $\mu$ is chosen properly. This region is the projection of the interval $0 < \phi < \mu$ of the Jordan frame potential (\ref{Vphi}) after the conformal transformation. One can verify that a successful slow-roll inflation in this part of the potential, can only be achieved with $\mu \gtrsim 10$. Sub-Planckian values fail to produce enough slow-roll inflation in this region. For instance, we present the results of our model with $\mu = 41.78~M_{\rm P}$ in fig. \ref{figure:rnsmu41p87chi4} . As noted before, the model with $\xi=0$, is still within the $95\%$ C. L. region of the  Planck 2018 data in the $r-n_{{}_S}$ plane. As $\xi$ increases,  $n_{{}_S}$ tends to decrease, while $r$ slightly increases a bit but decreases again, until the predictions of the model falls out of the $2\sigma$ region of the PLANCK 2018 data. We should recall that in the region $\chi_3<\chi<\chi_4$ only a finite number of e-folds are obtained, since spectator modes become tachyonic. Increasing the value of $\xi$, this interval shortens further and at some point the model cannot render sufficient number of e-folds anymore. In fig. \ref{figure:rnsmu41p87chi4}, we have plotted the number of $\xi$ at which only $60$ e-folds of inflation occurs. Incidentally the predictions of the model in this region is at the boundary of $1\sigma$ and $2\sigma$ viable regions of PLANCK 2018 data. As mentioned earlier, the finite number of e-folds can have some observational consequences \cite{Freivogel:2005vv}.

In region $\chi_2<\chi<\chi_3$ the same behavior is observed, although like the corresponding part in $\MM$-flation, this whole region suffers from runaway directions around the $SU(2)$ sector, as we will show in the next section.


\section{Isocurvature Spectra in non-$\MM$-flation} \label{section:isocurvature}

\subsection{Scalar Isocurvature Perturbations}\label{subsection:scalar-isocurvature}

Going from the Jordan to the Einstein frame, the potential for the scalar spectator fields will also change. In particular the Lagrangian of the scalar spectator fields after the conformal transformation becomes
\beq
\label{SEchiPsit}
 S_{\Psi}=\int d^{4}x\sqrt{-\tilde{g}}\left[ \frac{1}{2}\tilde{R}+\frac{1}{2}\left(\frac{d\chi}{d t}\right)^{2}-U(\chi)-\frac{1}{2}\frac{1}{\Omega^{4}} \sum_{i} \partial_{\mu} \Psi_{i} \partial^{\mu}\Psi_{i} -\tilde{V}_{(2)}\left(\chi,\Psi_{i}\right) \right]\,,
\end{equation}
where $U(\chi)$ is defined in eq. \eqref{UchiV0phi}, and
\begin{equation}
 \label{V2chiPsitildei}
\tilde{V}_{(2)}\left(\chi,\Psi_{i}\right)=\frac{V_{(2)}\left(\chi,\Psi_{i}\right)}{\Omega^{4}\left(\phi(\chi)\right)}=\frac{1}{2}\frac{M_{\Psi_{i}}^{2}\left(\phi(\chi)\right)}{\Omega^{4}\left(\phi(\chi)\right)}\Psi_{i}^{2}\,.
\end{equation}
One should also note that
\beq\label{kinetic-term-scalar-preheat}
\partial_{\mu} \Psi_{i} \partial^{\mu}\Psi_{i} =g^{\mu\alpha}\partial_{\alpha}\Psi_{i} \partial_{\mu} \Psi_{i} =\Omega^{2}\tilde{g}^{\mu\alpha}\partial_{\alpha}\Psi_{i} \partial_{\mu} \Psi_{i}\,.
\eeq
The summation over $i$ in \eqref{SEchiPsit} is over all the scalar spectator fields, whether they are $\alpha$-modes or $\beta$-modes. The potential in the Einstein frame is hence dependent on the potential of the Einstein frame by a positive conformal factor. That means that the region $\chi>\chi_4$, which is the projection of the region $\phi>\mu$ in $\MM$-flation, remains a local attractor. We will see that the same is true for the gauge modes in this region. We conclude that in the region $\chi>\chi_4$, the $SU(2)$ configuration remains a local attractor for non-$\MM$-flationary potential. Noting that the potential in this region of parameter space provides predictions that are compatible with the latest PLANCK bounds, this is indeed an appealing feature. In region $\chi_3<\chi<\chi_4$, due to the fact that the potential in Jordan frame became unstable in the $\omega=-3$ direction for values of $0.5\mu<\phi\leq 0.6144\mu$, the obtained number of e-folds one can obtain in this region of potential is indeed finite.  The left hilltop side, $\chi_2<\chi<\chi_3$, remains an unattractor again due to the tachyonic directions  in the spectator field space along $-79\leq \omega\leq -3$.

The spectator directions are frozen classically during inflation. However they can fluctuate quantum mechanically and their fluctuations will get imprinted on the CMB in the form of isocurvature perturbations, if they decay to radiation/dark matter. The kinetic term for $\Psi_i$, appeared in  \eqref{SEchiPsit}, is non-canonical. The formalism of cosmological perturbation theory in presence of such non-canonical two-field model has been developed in the \cite{Lalak:2007vi}. One can still define generalized curvature and entropy perturbations and derive the relevant equations of motion for the variables, as in the case of scalar fields with canonical fields \cite{Gordon:2000hv}. The trajectory along the $SU(2)$ direction is a straight line in the field space and there is no conversion of isocurvature to curvature perturbations, contrary to what happens in \cite{Ashoorioon:2008qr}. One can use the formalism of  \cite{Lalak:2007vi} and obtained the isocurvature power spectrum at the end of inflation numerically. The starting action in \cite{Lalak:2007vi} is of the form
\beq
S_{\Psi}=\int d^4 x \sqrt{-\tilde{g}}  \left(\frac{1}{2}\tilde{g}^{\mu\alpha}\partial_{\mu}\chi\partial_{\alpha} \chi-U(\chi)-\frac{e^{2 b(\chi)}}{2}\tilde{g}^{\mu\alpha}\partial_{\mu}\Psi_i\partial_{\alpha} \Psi_i - \tilde{V}_2(\chi,\Psi_i)\right)
\eeq
which can be matched with  \eqref{SEchiPsit}, if
\beq
b(\chi)=-\ln(\Omega(\phi(\chi)))\,.
\eeq
We used the code developed by K. Turzynski in \cite{Lalak:2007vi} to obtain the amplitude of perturbations. For $\mu\approx 0.01=M_{P}$ and $\xi\simeq 1591.91$ which yields $n_{{}_S}\simeq 0.9707$ and $r\simeq 0.0061$, the lightest $\alpha$ mode is $\omega=-3$ ($j=1$ $\alpha-$ mode), the amplitude of its power spectrum for the scales that exit the horizon $60$ e-folds before the end of inflation is $9.32\times 10^{-25}$. For $j=2$ and $j=3$ $\alpha$-modes, the amplitude of corresponding isocurvature spectra are, respectively $1.18\times 10^{-24}$ and $1.02\times 10^{-24}$.  Since the number of D3 branes needed in this case is about $N=16$, the largest $j$ $\alpha$-mode is $j=14$ for which the amplitude of corresponding isocurvature spectrum is $9.13\times 10^{-25}$. The amplitude of isocurvature spectra in this example turn out to be about $10^{15}$ smaller than the amplitude of density perturbations and they will be likely not going to be observable in the CMB. For $\beta$ modes, the lightest mode $j=1$, yields an even tinier value of isocurvature power spectrum, $\sim 3.91\times 10^{-78}$ at the CMB scales. Larger $j$ $\beta-$modes, yield even smaller amplitudes for their corresponding isocurvature spectra.

In the right hilltop region $\chi_3<\chi<\chi_4$, as noted before, the results are still inside the horizon for $\mu=41.87~M_P$ and $\xi=0$. The largest amplitude of isocurvature spectra, belongs to $j=1$ $\beta$ mode, which has an amplitude of $1.13\times 10^{-12}$. The degeneracy of this mode is 3. The next one in the tower of  $\beta$-modes is $j=2$, which has an amplitude of $8.8\times 10^{-18}$. It would depend on the scenario of reheating, but if there is conversion of all three $j=1$, $\beta$ isocurvature mode to curvature perturbations, there might be a chance of detection of these isocurvature modes of amplitude $\frac{P_{S_{\beta,1}}}{P_R}\sim 1.6\times 10^{-3}$ at the CMB scales.

\subsection{Gauge Isocurvature Perturbations}
\label{subsection:gauge-isocurvature-modes}

For the mass eignemodes coming from the gauge sector, the action in the Jordan frame is
\beqa\label{gauge-action}
S_{A}&=&\int d^{4}x\sqrt{-\tilde{g}}\left[ \frac{1}{2}\tilde{R}++\frac{1}{2}\left(\frac{d\chi}{d t}\right)^{2}-U(\chi)-\frac{1}{4\Omega^4(\phi(\chi))} \partial_{[\mu} A_{\nu]}   \partial^{[\mu} A^{\nu]}-\frac{V_2(\phi(\chi), A_{\mu})}{\Omega^4(\phi(\chi))} \right]\,,
\nonumber\\
&&\nonumber\\
\eeqa
where
\beq
V_2(\phi, A_{\mu}) =M_{A,j}^2(\phi(\chi))A^{\mu}A_{\mu}\,.
\eeq
We note that the metric in the gauge sector in action \eqref{gauge-action}, is still the metric of the Jordan frame. One can express the action completely in terms of the Einstein frame metric,
\beqa\label{gauge-action-2}
S_{A}&=&\int d^{4}x\sqrt{-\tilde{g}}\left[ \frac{1}{2}\tilde{R}++\frac{1}{2}\left(\frac{d\chi}{d t}\right)^{2}-U(\chi)-\frac{1}{4}\frac{\eta^{\mu\alpha}}{a^2} \frac{\eta^{\nu\beta}}{a^2} \partial_{[\mu} A_{\nu]}   \partial_{[\alpha} A_{\beta]} \right. \nonumber\\ && \left. - \frac{1}{2} \frac{\eta^{\mu\alpha}}{a^2} \frac{M_{A}^2(\chi) }{\Omega^2(\chi)}\eta^{\mu\alpha} A_{\mu} A_{\alpha}
\right]\,, \nonumber
\\
\eeqa
The disappearance of the factor of $\Omega$ and $a$ from the usual massless gauge field action is the reminder of the conformal invariance of the action. The mass term for the gauge field, nonetheless, gets a suppressing factor of $1/\Omega^2$ term. One can impose the gauge $A_0=\partial^{i}A_{i}=0$. The resulted equation of motion for the Fourier component of the gauge field components,
\begin{equation}
 \label{AmuAmuk}
 A_{i}(t,\mathbf{x})=\int\frac{d^{3}\mathbf{k}}{(2\pi)^{3/2}}\left[{A_i}_k(t)\hat{a}_{k}e^{-i\mathbf{k.x}}+{A_i}_k^{*}(t)\hat{a}_{k}^{\dagger}e^{i\mathbf{k.x}}\right].
\end{equation}
 with quantum number $j$, is
\beq\label{A-eq}
\ddot{A_i}_k+H \dot{A_i}_k+\left(\frac{k^2}{a^2}+\frac{M_{A,j}^2(\chi)}{\Omega^2(\chi)}\right) {A_i}_k=0\,,
\eeq
where $\dot{}\equiv \frac{d}{dt}$ and $H=\frac{\dot{a}}{a}$.

In order to find a harmonic oscillator form for the equation of motion of the gauge fields, we apply the rescaling
\begin{equation}
 \label{AbarikAik}
 \bar{A_i}_{k}\equiv a^{1/2}A_i{_k}\,.
\end{equation}
Therefore, from eq. (\ref{A-eq}) we obtain
\begin{equation}
 \label{Abarddotik}
 \ddot{\bar{A}}_{ik}+\omega_{\bar{A} k}^{2}\bar{A}_{i_k}=0\,,
\end{equation}
where,
\begin{equation}
 \label{omegaAk}
 \omega_{\bar{A}k}^{2}\equiv\frac{k^{2}}{a^{2}}+\frac{1}{4}H^{2}-\frac{\ddot{a}}{2a}+\frac{M_{A,j}^{2}(\chi)}{\Omega^{2}(\chi)}.
\end{equation}
To solve the differential equation (\ref{Abarddotik}), we impose the Bunch-Davies vacuum
\begin{equation}
 \label{AbarikBunchDavies}
 \bar{A}_{ik}\to\frac{1}{\sqrt{2\omega_{\bar{A} k}}}e^{-i\int^t \omega_{\bar{A} k} dt'},
\end{equation}
as the initial condition which is supposed to be valid when the modes are deep inside the Hubble horizon. Using the solution of the evolution equation \eqref{Abarddotik}, and noting that
\beq
B^i=g^{ij}\epsilon_{jlm} \partial_l A_m=\frac{\epsilon^{i}_{lm} k_l \bar{A}_m} {a^{5/2}}\,,
\eeq
 one can evaluate the spectrum of the magnetic fields which is given by
\begin{equation}
 \label{PB}
 \mathcal{P}_{B}=\frac{k^{5}}{2\pi^{2}}\frac{1}{a^{5}}\left|\bar{A}_{ik}\right|^{2}\,.
\end{equation}
The mass for the gauge mode $j=0$ remains zero and hence the $U(1)$ action keeps its conformal invariance. Due to this one would not expect an enhancement for this mode during inflation.
For $\mu = 0.01, \, \xi = 500$ the amplitude of the power spectra for the modes that exit the horizon about 60 e-folds before the end of inflation are $\mathcal{P}_{B}^{j=0}\simeq 9.52\times 10^{-130}$ and $\mathcal{P}_{B}^{j=1}\simeq 4.85 \times 10^{-158}$, which are quite small. For $\mu = 0.01, \, \xi = 1520$, the values of these spectra are, respectively, $2.68\times 10^{-131}$ and $7.44\times 10^{-159}$.


\section{Preheating in non-$\mathbb{M}$-flation}
\label{section:preheating}

\begin{figure}
\centering
\includegraphics[scale=0.36]{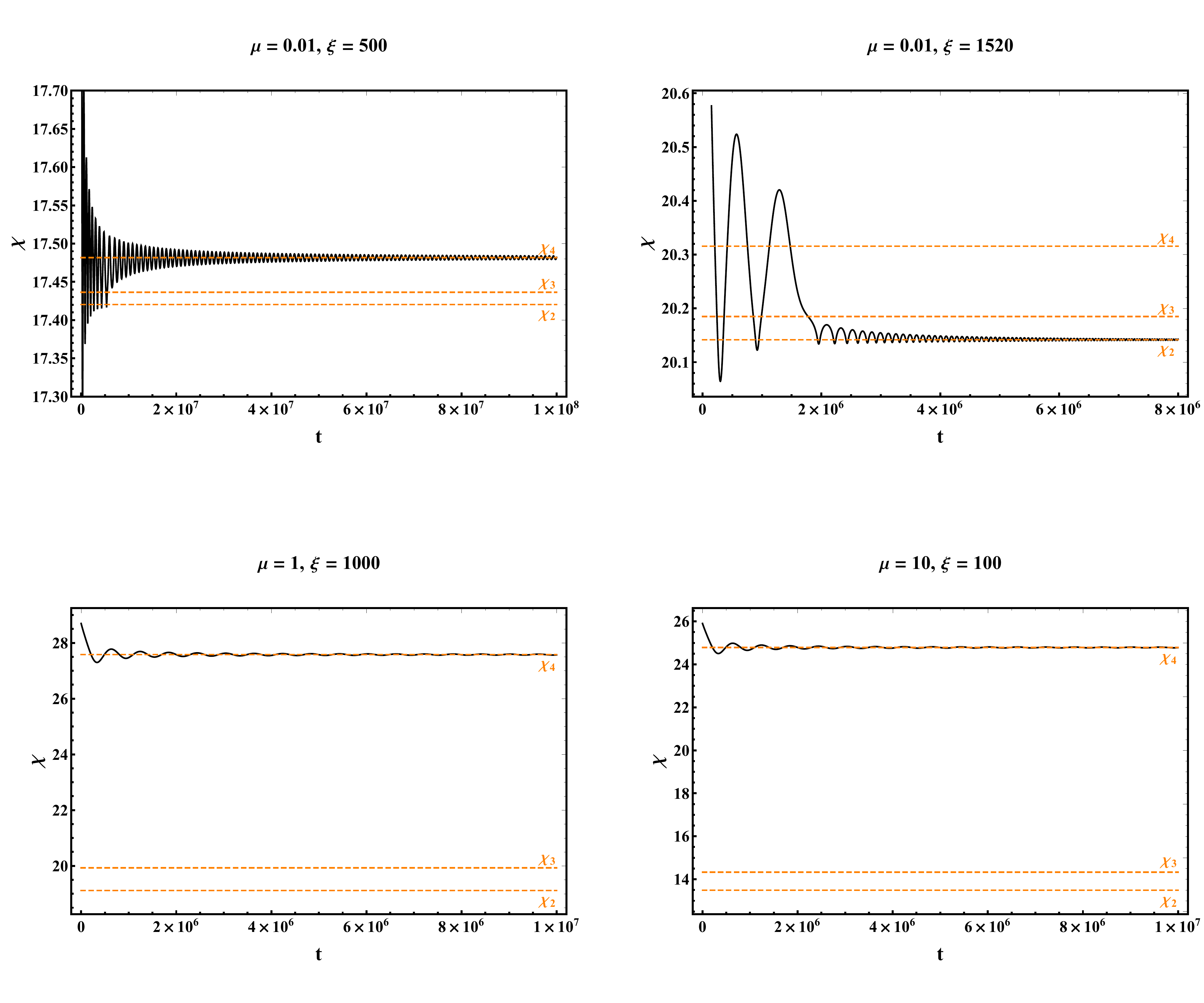}
 \caption{
 The time evolution of the inflaton field in the Einstein frame, $\chi$, for given parameters $\mu$ and $\xi$. The orange lines in the figure refer to the minima placed at $\chi_2$ and $\chi_4$, and also the maximum of the bump at $\chi_3$.
 }
 \label{figure:chit-mu-less-mpl}
\end{figure}

During preheating, it is presumed that some extra matter, dubbed as preheat fields which are coupled to the inflaton, start depleting its energy after inflation terminates. This draining from the inflaton field can happen either perturbatively \cite{Dolgov:1982th, Abbott:1982hn}, or non-perturbatively when the inflaton oscillates around its minimum \cite{Traschen:1990sw, Kofman:1994rk, Shtanov:1994ce, Kofman:1997yn}. The decay of the inflaton to the preheat fields can warm up the universe again. The setup of Matrix Inflation is in general equipped with spectator modes whose masses depend on the value of the inflaton field and thus can act as such preheat fields at the end of inflation\footnote{In $\lambda \phi^4$ theory that can be realized in ungauged $\MM$-flation \cite{Ashoorioon:2009wa}, we showed that the amount of depletion of the energy of the inflaton by these preheat fields is negligible during the slow-roll phase of M-flation with $H \sim 10^{-5}$, despite the large degeneracy of the spectator modes.}. Unfortunately in $\MM$-flation these spectators fail to reheat the universe when inflation ends in the spontaneously symmetry-breaking vacuum, $\phi=\mu$ \cite{Ashoorioon:2013oha}. The $j=0$ gauge mode remains decoupled from the inflaton, although it remains massless and hence lighter than the inflaton around $\phi=\mu$. The next  massive mode in the tower of preheat modes is $j=1$ $\beta$-modes which has a mass equal to the inflaton at the symmetry-breaking vacuum.  This precludes the decay of the inflaton to this mode, even though it has a cubic interaction with the inflaton. The larger $j$ modes are heavier than the inflaton and the perturbative decay of the inflaton to them is a priori impossible. As it was shown in \cite{Ashoorioon:2013oha}, {\it narrow} resonance band is possible for $1\leq j\leq 3$ $\beta$-mode and $j=1$ $\alpha$-mode. In $\MM$-flation though, considering the expansion of the universe, the  $k$-mode falls out of the resonance band and parameteric resonance is not successful.

Although one can in principle try to construct the (Beyond) Standard Model sector locally in another part of the compactified manifold and make it coupled to the $\MM$-flationary sector, it is unfortunate that in Matrix inflation, the spectator fields fail to preheat the universe, even though they have some of the basic ingredients for realization of the phenomenon. Below we will show that in the non-$\MM$-flation scenario, the kinematics and dynamics of the model is such that one can have a successful partial or complete preheating around the symmetry-breaking vacuum.  As we will see in some cases that $\mu$ is sub-Planckian, the inflaton rolls over the barrier that separates the symmetric and symmetry-breaking vacuum around which some of the modes enjoy the tachyonic instability and therefore there will be an explosive production of spectators modes in such cases. As before, we will  treat the scalar spectator and gauge spectator modes separately below.

\subsection{Scalar preheat fields}
\label{subsection:scalar-preheat-fields}

We first concentrate on the scalar preheat fields $\Psi_i$. The action for the scalar spectator modes that are now supposed to play the role of preheat fields is given by \eqref{SEchiPsit}, where one should consider that the kinetic term is non-minimal after considering the eq. \eqref{kinetic-term-scalar-preheat},
\beq
\label{SEchiPsit2}
 S_{\Psi}=\int d^{4}x\sqrt{-\tilde{g}}\left[ \frac{1}{2}\tilde{R}+\frac{1}{2}\left(\frac{d\chi}{d t}\right)^{2}-U(\chi)-\frac{1}{2}\frac{1}{\Omega^{2}(\chi)} \sum_{i} \tilde{g}^{\mu\alpha} \partial_{\mu} \Psi_{i} \partial_{\alpha}\Psi_{i} -\tilde{V}_{(2)}\left(\chi,\Psi_{i}\right) \right]\,.
\end{equation}
The kinetic term for the preheat fields could be brought to the canonical form, if one introduces the new variable
\beq\label{psitilde}
\tilde{\Psi}_{i}=\Omega(\chi) \Psi_{i}\,.
\eeq
However this change of variable will induce a kinetic mixing between the inflaton field, $\chi$, and the new preheat field, $\tilde{\Psi}_{i}$. It also induces a correction to the interacting potential proportional to the kinetic term of the inflaton field $\dot{\chi}^2$. The action takes the form in terms of the new variable, $\tilde{\Psi}_{i}$,
\begin{align}
 S_{\tilde{\Psi}}= &  \int d^{4}x\sqrt{-\tilde{g}}\left[  \frac{1}{2}\tilde{R}+ \frac{1}{2}\left(\frac{d\chi}{d t}\right)^{2}-U(\chi)-\frac{1}{2}\sum_{i} \tilde{g}^{\mu\alpha} \partial_{\mu} \tilde{\Psi}_{i} \partial_{\alpha}\tilde{\Psi}_{i} -\bar{V}_{(2)}\left(\chi,\tilde{\Psi}_{i}\right)\right.
 \nonumber \\
  & \left.  + \frac{d\ln \Omega(\chi)}{d\chi} \frac{d\chi}{d t }\frac{d\tilde{\Psi}_i}{d t}\right]\,,
\end{align}
where the new potential, $\bar{V}_{(2)}$, has also obtained some new corrections proportional to the kinetic term of the inflaton, $\chi$, as
\begin{equation}
 \label{V2chiPsitildei2}
\bar{V}_{(2)}\left(\chi,\tilde{\Psi}_{i}\right)=\frac{1}{2\Omega^{2}(\chi)}\left[M_{\tilde{\Psi}_i}^{2}+\left(\frac{d\Omega}{d \chi}\right)^{2} \left(\frac{d\chi}{d t}\right)^{2} \right]{\tilde{\Psi}}_{i}^{2}\,.
\end{equation}
We now perform the conventional second quantization and Fourier decomposition,
\begin{equation}
 \label{PsiiPsiik}
\tilde{\Psi}_{i}(t,\mathbf{x})=\int\frac{d^{3}\mathbf{k}}{(2\pi)^{3/2}}\left[\tilde{\Psi}_{i_k}(t)\hat{a}_{k}e^{-i\mathbf{k.x}}+\tilde{\Psi}_{i_k}^{\ast}(t)\hat{a}_{k}^{\dagger}e^{i\mathbf{k.x}}\right].
\end{equation}
We also introduce the rescaled field $\bar{\Psi}_{i_k}\equiv a^{3/2}\tilde{\Psi}_{i_k}$, to get rid of the Hubble friction term in the equation of motion for $\tilde{\Psi}_{i\,k}$. The evolution equation for the new variable takes a harmonic oscillator form,
\begin{equation}
 \label{d2Psihatikdttilde2}
 \frac{d^{2}\bar{\Psi}_{i_k}}{d t ^{2}}+\omega_{\Psi_k}^{2}\bar{\Psi}_{i_k}=0\,,
\end{equation}
where,
\begin{equation}
\omega_{\bar{\Psi} k}^{2}\equiv\frac{k^{2}}{a^{2}}+\frac{M_{\Psi}^{2}}{\Omega^{2}}-\frac{3}{4}H^{2}-\frac{3}{2}\frac{\ddot{a}}{a}+3H\frac{\dot{\Omega}}{\Omega}-2\left(\frac{\dot{\Omega}}{\Omega}\right)^{2}+\frac{\ddot{\Omega}}{\Omega}
  \label{omegaPsik}
\end{equation}
To solve the differential equation (\ref{d2Psihatikdttilde2}), we impose the Bunch-Davies vacuum for the modes at the onset of preheating,
\begin{equation}
 \label{PsihatikBunchDavies}
 \bar{\Psi}_{i_k}\to\frac{1}{\sqrt{2\omega_{{}_{\bar{\Psi}_k}}}}e^{-i \int \omega_{{}_{\bar{\Psi}_k}} d t^{\prime}}\,.
\end{equation}
The solution of the differential equation (\ref{d2Psihatikdttilde2}) is used to calculate the number density of the produce particles in mode $k$,
\begin{equation}
 \label{nPsik}
 n_{\bar{\Psi}_k}=\frac{\omega_{\bar{\Psi}_k}}{2}\left[\frac{1}{\omega_{\Psi_k}^{2}}\left|\frac{d\bar{\Psi}_{i_k}}{d t}\right|^{2}+\left|\bar{\Psi}_{i_k}\right|^{2}\right]-\frac{1}{2}\,.
\end{equation}

\begin{figure}
 \centering
 \includegraphics[scale=0.2]{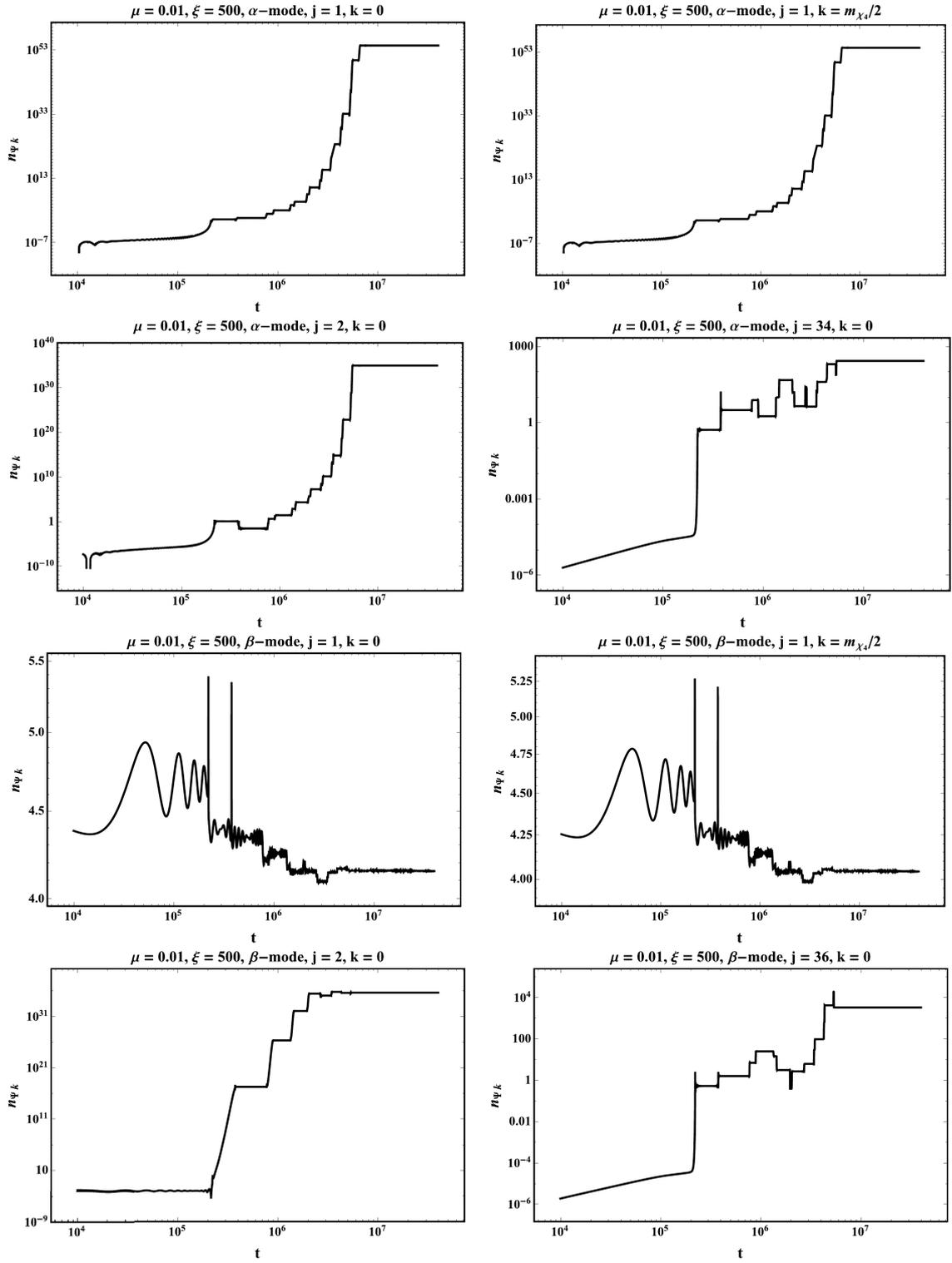}
 \caption{The time evolution of the logarithm of the number density of the various scalar preheat fields, $\Psi_i$, vs. logarithm of time, for $\mu = 0.01$ and $\xi = 500$.}
 \label{figure:nPsiktmu0p01xi500}
\end{figure}

To compute the number density using the above equation, the equation of motion for the preheat fields, $\bar{\Psi}_{i_k}$, eq. \eqref{d2Psihatikdttilde2} needs to be solved along with the background equation of motion for $\chi$. Depending on the values of $\mu$ and $\xi$ various scenarios can arise. For sub-Planckian values of  $\mu$, the $\chi$ field may roll over the bump between the symmetry-breaking and symmetric vacua, $\chi_4$ and $\chi_2$ before settling around one of the vacua. As we noted before, $\chi_2$ is the vacuum that corresponds to $\phi=0$ in the Jordan frame and some of the spectator fields remain light or tachyonic around this vacuum. This will allow for an explosive particle production of the preheat fields and full draining of the inflaton energy. For example for $\xi=500$ and $\mu=0.01~M_{P}$ the inflaton, $\chi$, passes through the symmetric vacuum, $\chi_2$, for few oscillations before getting trapped in the valley around $\chi_4$, please see the left upper graph in fig. \ref{figure:chit-mu-less-mpl}. For $\xi=1520$ and $\mu=0.01~M_{P}$, the inflaton $\chi$, ends up oscillating around the symmetric vacuum, $\chi=\chi_2$, please see the upper right graph in fig. \ref{figure:chit-mu-less-mpl}. For values of $\mu\gtrsim M_{P}$, and different values of $\chi$, the inflaton does not seem to be able to pass through the barrier that separates the two vacua from each other, please see the lower two graphs in fig. \ref{figure:chit-mu-less-mpl}. We will analyze each of these cases separately below.

Before closing this subsection, it is useful to compute the effective mass of the inflaton field, $\chi$, in the Einstein frame around each of the two vacua,  $m_{\chi}^2\equiv\partial^{2}U(\chi)/\partial\chi^{2}$. One can show that the effective mass around the minima $\chi_2$ and $\chi_4$ are respectively given by
\begin{align}
 \label{mchi2}
 m_{\chi_{2}}^{2} & =\frac{\lambda_{{\rm eff}}\mu^{2}}{2}\,,
 \\
 \label{mchi4}
 m_{\chi_{4}}^{2} & =\frac{\lambda_{{\rm eff}}\mu^{2} M_{P}^2}{2\left[\xi(6\xi+1)\mu^{2}+M_{P}^2\right]}\,.
\end{align}
As the above formulae suggest, the mass of the inflaton around the symmetry-breaking vacuum in the Einstein frame gets suppressed by a factor of $\left[1+\xi (6\xi+1)\mu^2/M_{P}^2\right]^{-1/2}$ with respect to its mass in the Jordan frame around the same vacuum. The mass of the inflaton field around the symmetric vacuum remains the same as the mass of the inflaton field in the Jordan frame around the same vacuum. Having the time evolution of $\chi$ in hand, we turn proceed to computing the number density of the scalar preheat fields after the termination of inflation.
\begin{itemize}
\item  $\mu=0.01~M_{P}$ and $\xi=500$: As noted above, with such values for the symmetry breaking vev, the inflaton interpolates between two vacua before settling for the symmetry-breaking vacuum. We have plotted the number density of $\alpha_{1}$ and $\beta_{ 1}$ scalar preheat modes for $k=0$ and $k=\frac{m_{\chi_4}}{2}$, please see fig. \ref{figure:nPsiktmu0p01xi500}. In all the cases the number density of the particles grows stochastically. The growth is faster when the inflaton samples the symmetric vacuum, as the preheat mode becomes tachyonic around that mode for a small interval of time. For $k=0$, the growth almost shuts off when the inflaton starts solely oscillating around the symmetry-breaking. Similar behavior and amplitude growth in the number density of particles is observed  for $k=\frac{m_{\chi_4}}{2}$, so the resonance band seems to be quite  broad. Similar behavior for $k=\frac{m_{\chi_2}}{2}$ is observed was noticed although we have not demonstrated the plot to prohibit the cluttering of the paper. As one increases $j$, the produced number of particles settles on a smaller value, as we have demonstrated the results for $j=2$ and $j=34$. For $j=1$, $\beta$-mode and both $k=0$ and $k=\frac{m_{\chi_4}}{2}$, on the other hand, the growth of the particles is much suppressed, as no tachyonic mass for this preheat mode develops around the symmetric vacuum for this particular mode, please see fig. \ref{figure:nPsiktmu0p01xi500}. As one steps up to $j=2$, the number density of the corresponding mode grows. With increasing $j$, the number growth of particles become further suppressed since the modes become heavier. For $\mu=0.01~M_{P}$ and $\xi=500$, the number of D3 branes needed to reduce the undressed coupling to the one required by observation, $\lambda_{\rm eff}=1.5987\times 10^{-4}$ is about $N\approx 36$. This means that  for $\alpha$-modes, $j_{\rm max}=34$ ($\omega=-36$) and for $\beta$-modes, $j_{\rm max}=36$ ($\omega=35$).

\begin{figure}
 \centering
  \includegraphics[scale=0.2]{nPsiktmu0p01xi1520}
 \caption{The time evolution of the logarithm of the number density of various scalar preheat fields, $\bar{\Psi}_i$ as a function of time, for $\mu = 0.01$ and $\xi = 1520$.}
 \label{figure:nPsiktmu0p01xi1520}
\end{figure}

 \begin{figure}
 \centering
\includegraphics[scale=0.2]{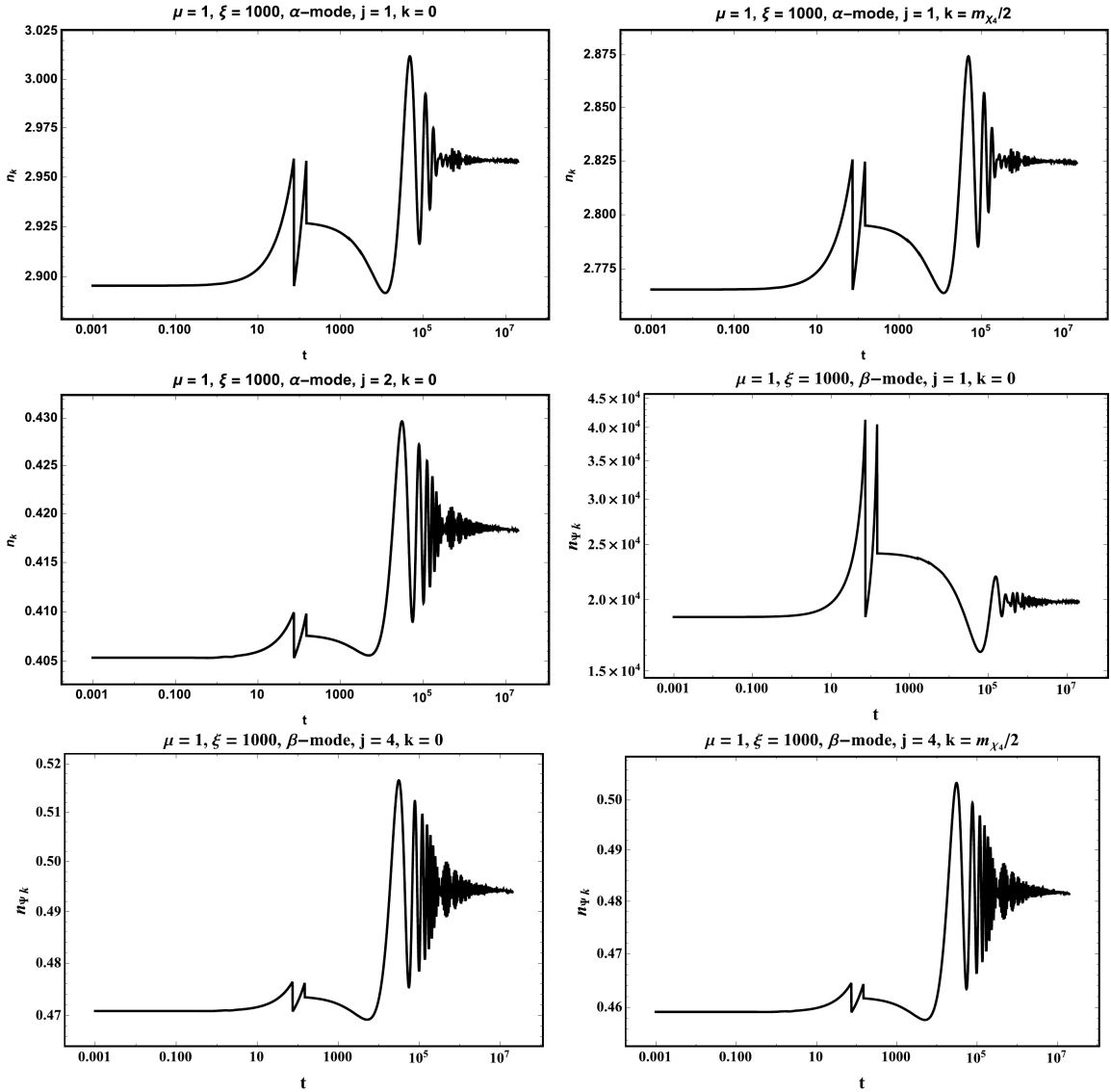}
\caption{The time evolution of number density of the scalar preheat fields $\Psi_i$ for $\mu = 1$ and $\xi = 1000$.}
 \label{figure:nPsiktmu1xi1000}
\end{figure}

 \item  $\mu=0.01~M_{P}$ and $\xi=1520$: In this case the inflaton interpolates between two vacua until it finally gets trapped around the symmetric vacuum. For $j=1$ $\alpha$ mode the particle production is explosive due to the narrow tachyonic region around $\chi_2$. As one enhances $j$ for the $\alpha$ mode becomes slower than $j=1$ but still it is explosive. There is no stochastic particle production $j=1$ $\beta$ mode particles on the other hand, please see fig. \ref{figure:nPsiktmu0p01xi1520}. However, once again with increasing $j$ for the $\beta$ modes, the growth of number density of particles  becomes faster.

\begin{figure}
 \centering
\includegraphics[scale=0.2]{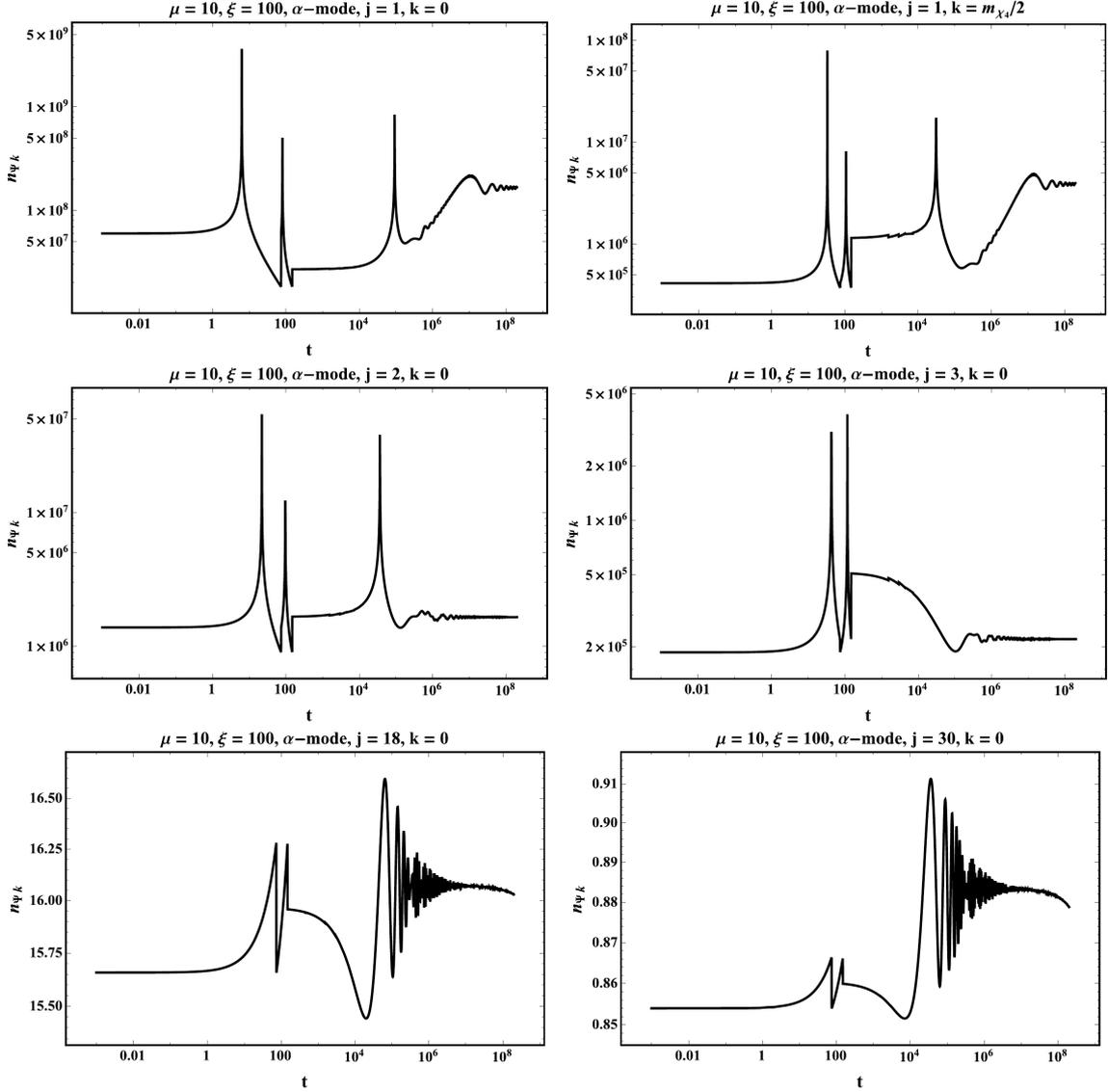}
 \caption{The time evolution of number density for some of the $\alpha$-preheat modes for various values of $k$ for $\mu = 10$ and $\xi = 100$.}
 \label{figure:nPsiktmu10xi100-alpha}
\end{figure}

 \item  $\mu=1~M_{P}$ and $\xi=1000$: In this case, the inflaton only oscillates around the symmetry-breaking vacuum. Number of the D3 branes which is needed to suppress the bare coupling of order one to the required observed value will be $\sim 18$. For $j=1$ $\alpha$ preheat mode, although stochastic preheating occurs to some extent, neither for $k=0$ nor for $k=\frac{m_{\chi_4}}{2}$, the final number density, $n_{\bar{\Psi_i}}$, increases that much, please see fig. \ref{figure:nPsiktmu1xi1000}. This is despite the fact that $n_{\bar{\Psi_i}}$ already has an enhancement factor roughly proportional to $(1+\xi\mu^2)$ due to the definition of new variable $\tilde{\Psi}$ in eq. \eqref{psitilde}. Despite all this the number density of the produced particles only get enhanced to values of order $n_{\bar{\Psi}_{\alpha,1}}\approx 2.82-2.95$, aside from some spiky features believes to be the result of the sudden change of behavior in the variation of the potential and the kinetic term \cite{Ema:2016dny}. This is also because for $j=1$, one can still have narrow resonance stochastic preheating. For larger values of $j$ $\alpha$-preheat modes, the produced number of particles settles down on a smaller value and the produced number of particles decreases exponentially with time, although with a very small slope, which shows that for them it becomes impossible to keep the momentum $k$ in the instability band. For $j=3$ $\alpha$-mode the produced number of particles reaches up to $\sim 0.094$, which decays almost linearly later as time passes by. On the other hand, $j=1$ $\beta$ preheat mode settles on a much larger value of number density for particle for $k=\frac{m_{\chi_4}}{2}$, $n_k^{\rm f}\sim 9700$, please see fig. \ref{figure:nPsiktmu1xi1000}. With increasing $j$, the final value for the number density of particles gets suppressed further. For $j=5$, the number density of particles reaches values of order $0.1$, which like the large $j$ $\alpha$-modes decays quasi-exponentially with a negative exponent from there, again due to given momentum $k$ not being able to remain in the instability band of the narrow resonance stochastic particle production. We conclude that in this case the efficient preheat mode is the $j=1$ $\beta$. Please note that even though the obtained number density of the produced particles is not that large, this should be enough to transfer the energy of the inflaton completely to  $j=1$ $\beta$.  Noting that the resonance band is broad enough, $\Delta k\sim \frac{m_{\chi_4}}{2}$, one can estimate the energy density taken away by this preheat mode solely to be about $10^4 m_{\chi_4}^4\sim 10^{-16}~M_P^4$.  Although this is almost six orders of magnitude smaller than the energy of the inflaton at the end of inflation, still the resulted reheating temperature after thermalization of this amount of energy is not only more than enough to satisfy the nucleosynthesis constraint $T_{\rm RH}\gtrsim 10^{-3}$ GeV, but also above the electroweak phase transition temperature. We leave a more exact treatment of the reheating process and the aftermath to a future investigation.

 \begin{figure}
 \centering
\includegraphics[scale=0.2]{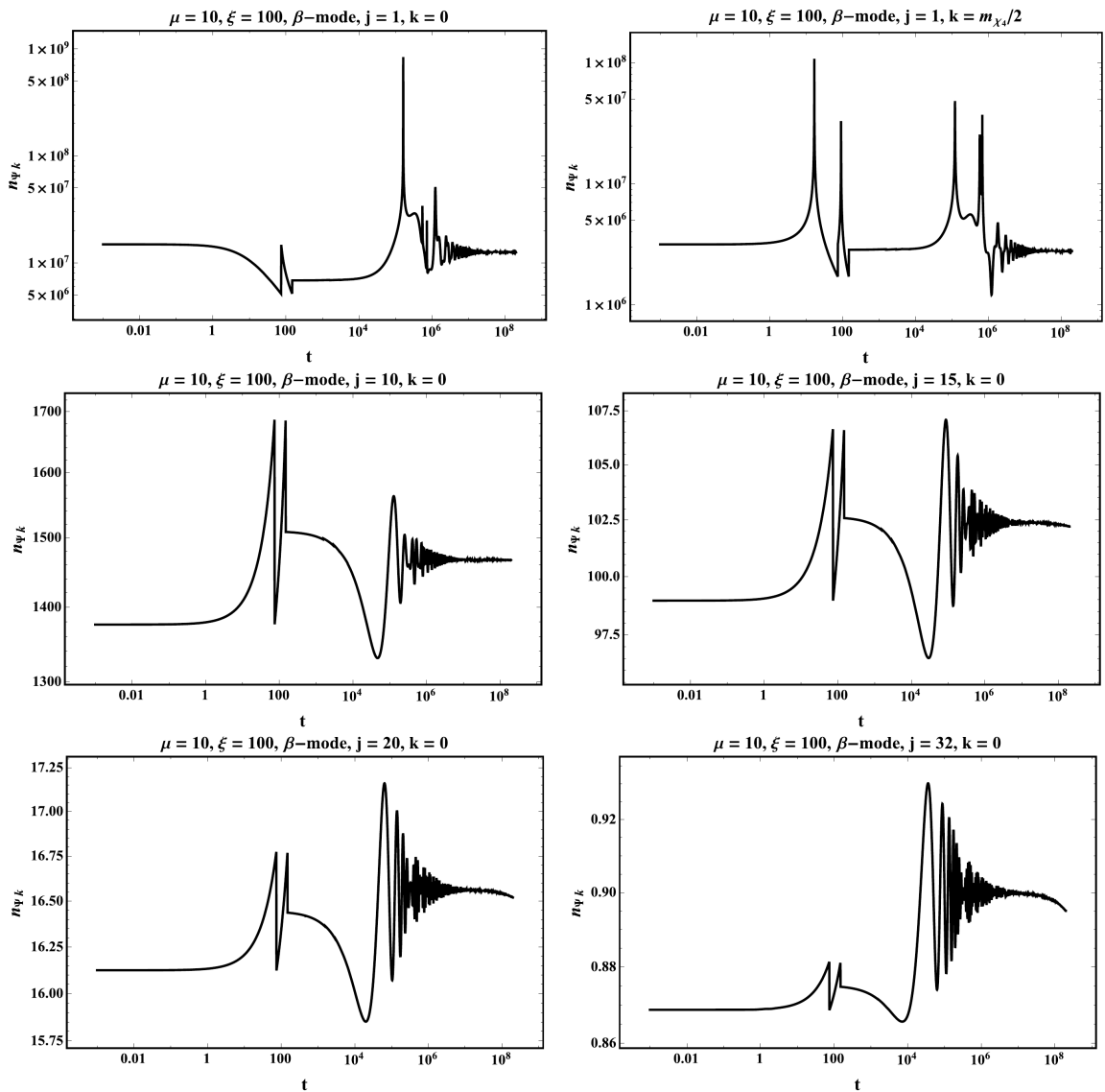}
\caption{The time evolution of number density for some of the $\beta$ preheat modes for various values of $k$ for $\mu = 10$ and $\xi = 100$.}
 \label{figure:nPsiktmu10xi100-beta}
\end{figure}

  \item  $\mu=10~M_{P}$ and $\xi=100$: The required number of D3 branes is about $N\sim 84$ in this case. A behavior similar to the case of $\mu=1~M_{P}$ and $\xi=1000$ is observed, with this discrepancy that $\alpha$ modes in this case are more effective in preheating in comparison with the $\beta$ modes. For $j=1$ $\alpha$-mode the number density of particles settle on the value of $\sim 1.7\times 10^8$ for $k=0$ and $\sim 4\times 10^6$ for $k=\frac{m_{\chi_4}}{2}$ for large values of time, which is much larger than the previous case studies with Planckian symmetry-breaking vacuum. We have also plotted the log-log plot of number density vs. time to see the transient spikes the number density of the produced particles as a function of time. As it has been pointed out in \cite{Tsujikawa:1999jh, DeCross:2015uza, Ema:2016dny}, these spikes are due to spiky behavior of the effective inflaton's mass in the Jordan frame. As $j$ increases, the number of produced particles the stochastic process produce gets suppressed. We have plotted them for $j=2,\,3,\, 18,\,30$ for $k=0$, please see fig. \ref{figure:nPsiktmu10xi100-alpha}. For $j=30$ , the final number of particles drops below one. Larger values of $j$, settle on a smaller number density.  The other thing one should note is that as one enhances the $j$ number, once $n_{\bar{\Psi}}$ reaches its maximum value, it starts to decrease very slowly with time, which is indicative of the fact that as the bare mass of the preheat mode increases with $j$, the narrow resonance band fails to keep up with the expansion of the universe. Similar pattern of particle production is observed for the tower of $\beta$-modes as preheating channels, please see figure \ref{figure:nPsiktmu10xi100-beta}, For $j=1$ $\beta$ mode, the number density of particles reaches values of order $n_{\beta_1}=1.3\times 10^7$ which is smaller than the corresponding value for $j=1$ $\alpha-$mode by almost an order of magnitude. It is for $j\gtrsim 32$ where the number density of the particles fall below one for $k=0$.

\item $\mu=41.87~M_{P}$ and $\xi=3.638\times 10^{-4}$: Not only for values of $\xi\gg 1$ and in the region $\chi_3<\chi<\chi_4$, the non-minimal coupling helps in facilitating the preheating. The values for the parameters in this case, correspond to what would keep the predictions of the hilltop region in the 2$\sigma$ region of the PLANCK 2018 data and yield 60 e-folds of inflation.  We have plotted the number density of particles for $j=1$ $\alpha$ mode in a log-log plot. As it can be seen contrary to the case where $\xi=0$, in this case we have a successful stochastic particle production during reheating. Nonetheless the final number of produced particles, is not that large, $n_{\alpha,1}^{f}\sim 470$. For $k=\frac{m_{\chi_4}}{2}$, the final number of produced particles is smaller, $n_{\alpha,1}^{f}\sim 320$. We conclude that in these cases the stochastic resonance happens, it occurs very feebly and lead to smaller reheating temperature after the energy of the excited modes is homogenized by interactions. In cases with small $\xi$, however, the required number of D3 branes remains of order ${\rm few}\times 10^4$.
\end{itemize}

\begin{figure}
 \centering
 \includegraphics[scale=0.7]{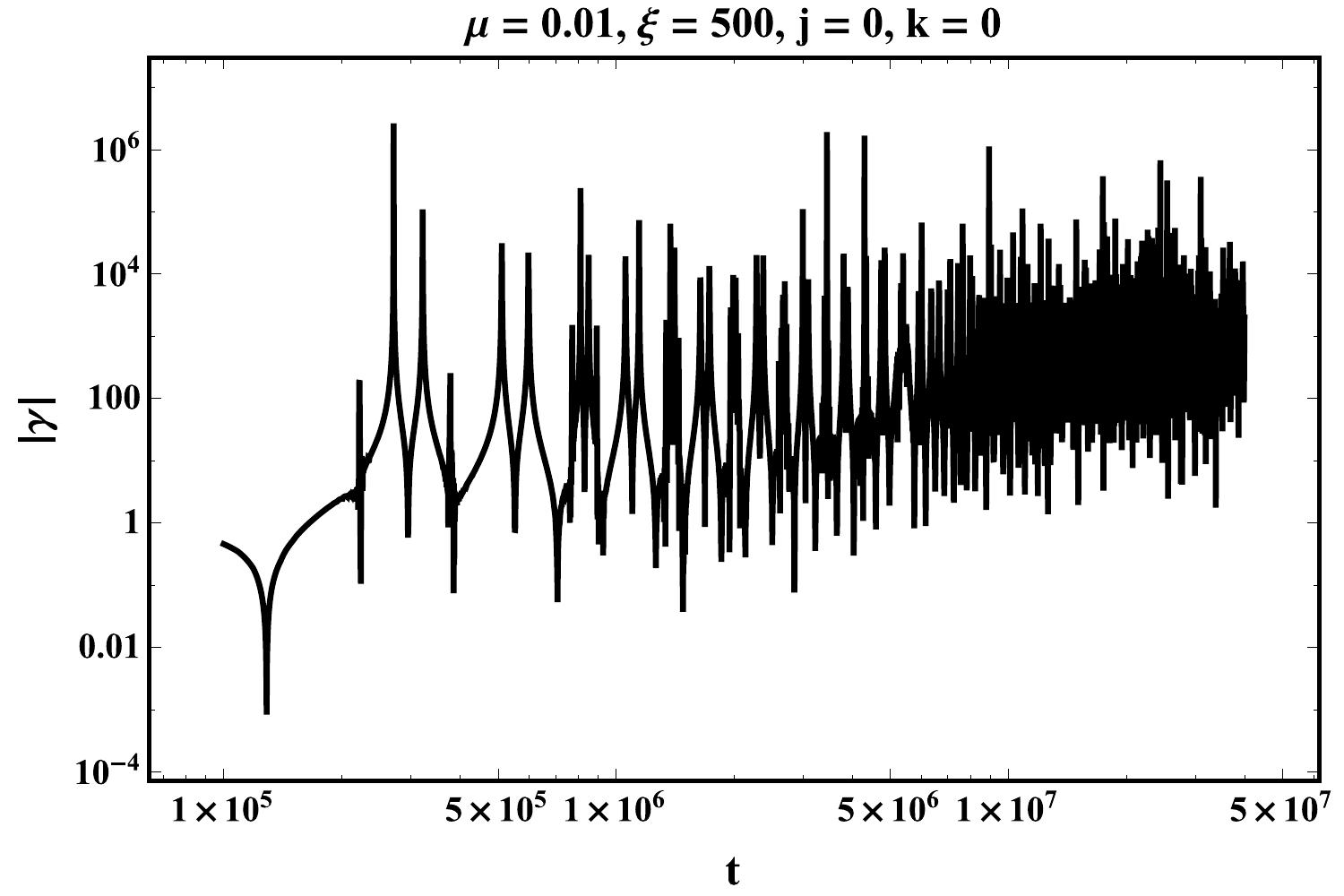}
 \caption{The time evolution of the adiabaticity parameter.}
 \label{figure:gamma}
\end{figure}

 \begin{figure}
 \centering
\includegraphics[scale=0.2]{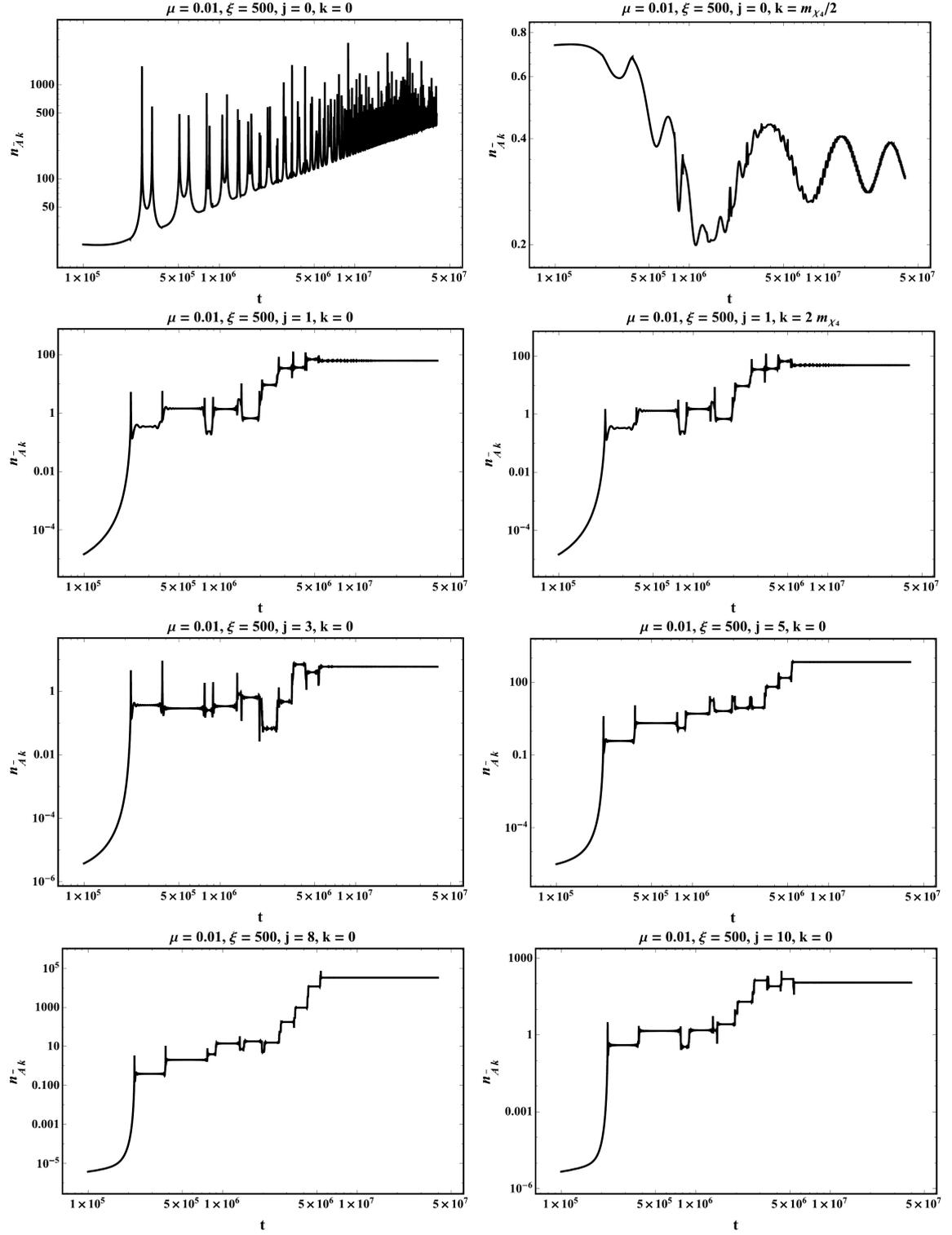}
\caption{The time evolution of number density for the gauge preheat modes, $n_{\bar{A} k}$ for various values of $j$ and $k$ for $\mu = 0.01~M_{P}$ and $\xi = 500$. }
 \label{figure:ngtmu0p01xi500}
\end{figure}


\subsection{Gauge preheat fields}
\label{subsection:gauge-preheat-fields}

In this subsection, we investigate the preheating effects of the gauge modes in non-$\mathbb{M}$-flation model. In subsection \ref{subsection:gauge-isocurvature-modes}, we argued that the gauge fields behaves as the isocurvature modes during inflation but the observational consequences of their perturbations is negligible in comparison with the inflaton field perturbations. In this subsection we check if the gauge fields play any effective role in the preheating process after inflation. To determine the evolution of the gauge fields we still use the equation of motion (\ref{Abarddotik}) with the initial condition (\ref{AbarikBunchDavies}). Applying the solution of the differential equation (\ref{Abarddotik}), we can compute the number density of the gauge fields given by
\begin{equation}
 \label{nAk}
 n_{\bar{A} k}=\frac{\omega_{\bar{A} k}}{2}\left(\frac{1}{\omega_{\bar{A} k}^{2}}\left|\dot{\bar{A}}_{i k}\right|^{2}+\left|\bar{A}_{i k}\right|^{2}\right)-\frac{1}{2}\,,
\end{equation}
where
${\omega}_{\bar{A} k}^{2}$ is given in eq. \eqref{omegaAk}. We have used this equation to plot evolution of the number density of gauge modes in the non-$\mathbb{M}$-flation model. Below, we will analyze each of the five examples discussed in the previous section separately.

\begin{itemize}

\begin{figure}
 \centering
 \includegraphics[scale=0.2]{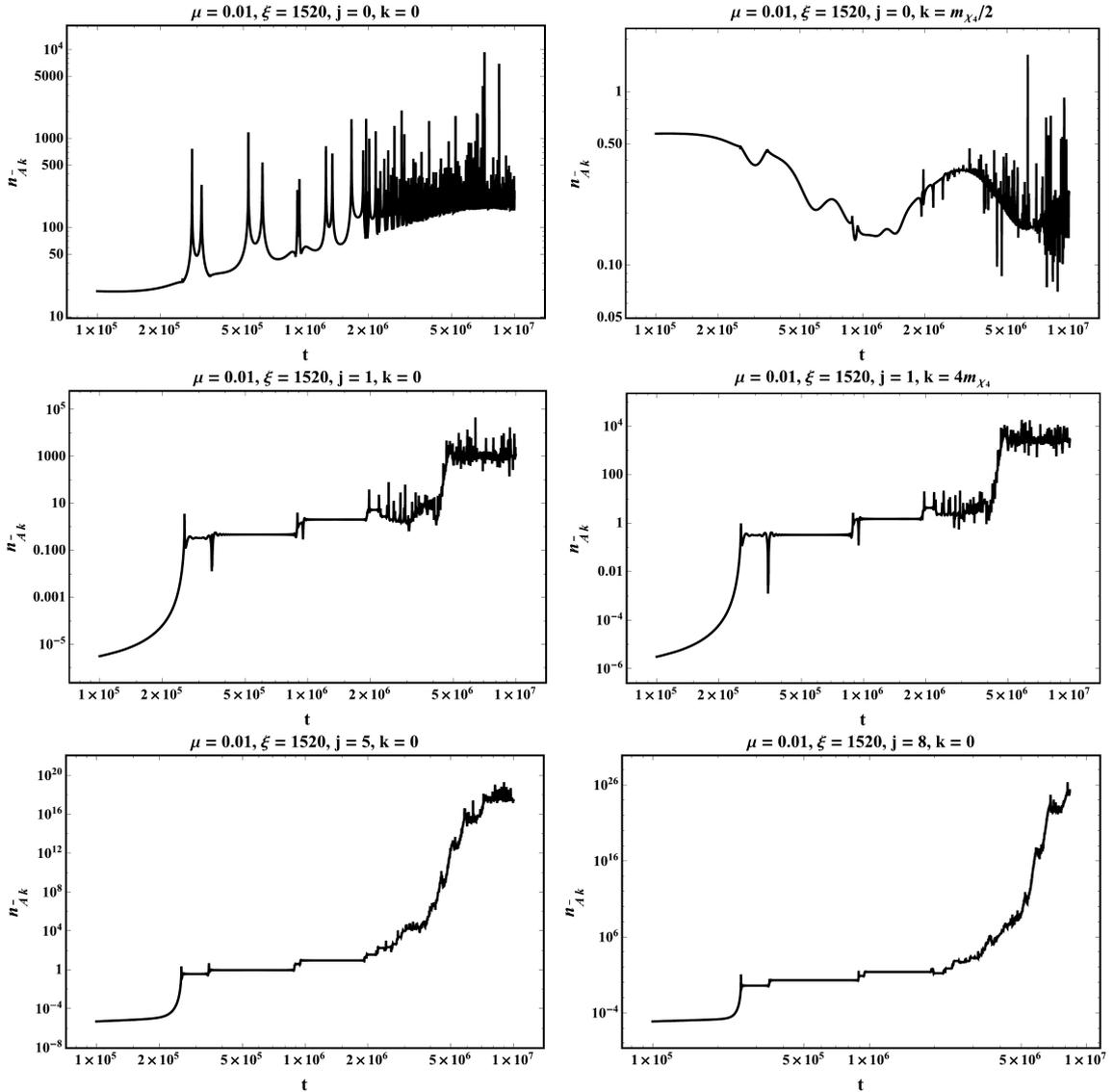}
 \caption{The time evolution of number density of the gauge preheat fields $\bar{A}_i$ for $\mu = 0.01$ and $\xi = 1520$.}
 \label{figure:ngtmu0p01xi1520}
\end{figure}

\item $\mu=0.01~M_{P}$ and $\xi=500$: The results has been presented in fig. \ref{figure:ngtmu0p01xi500}. For $j=0$, one can see spikes in $n_{\bar{A} k}$ for k=0 and $j=0$, which corresponds to the U(1) sector of U(N) gauge group. Such spikes in the behavior of the U(1) gauge field appears despite the independence of the mass of the U(1) gauge field from the value of inflaton for $j=0$. One can have a look at the non-adiabaticity parameter,
\beq
\gamma\equiv \frac{{\dot{\omega}}_{\bar{A} k}}{\omega_{\bar{A} k}^2}\,,
\eeq
and see that indeed it shows some spiky behavior at which it becomes quite larger than one. In particular whenever the inflaton passes through the vacua in the Jordan frame, $\chi_2$ and $\chi_4$, one notices a spike in the number density of particles, please see fig. \ref{figure:gamma}. If this U(1) field is identified with the Standard Model weak hypercharge, such electromagnetic fields {\it may} will correspond to the usual (electro)magnetic fields. However such electromagnetic fields have a too small wavelength to explain the observed intragalactic magnetic fields observed today \cite{Kronberg:1993vk,Bonafede:2010xg}. The width of the resonance band for $j=0$ is quite small. We have plotted the number density as a function of time for $k=m_{\chi_4}/2$ and it can be observed that the amplitude of the spikes get suppressed substantially.  For $j=1$ and $k=0$ besides the spiky behavior that appears when the mode crosses the two minima in the Einstein frame potential, one notices that  $n_{\bar{A} k}$, in between, roughly reaches constant values. Contrary to $j=0$, the width of the resonance band is much larger and in fact reaches about half its amplitude, only about $k\approx 3~m_{\chi_4}$.  Increasing $j$ from 1 to 4, initially one notices that the particle production becomes more suppressed, but from $j=5$ to larger values of $j$, the number density starts to soar up again. This is attributed to the fact that for $2\leq j\leq 4$, even though the inflaton passes through the symmetric vacuum, $\chi_2$, the amount of adiabaticity violation for such small $j$'s is still small. Also, the mass of these gauge fields around the symmetry-breaking vacuum starts to increase with $j$ in this range and therefore they become more costly to get produced. With increasing $j$, the former effect takes over the latter and the number density of the corresponding $j$ vector mode starts to increase, until it is again dominated by the latter effect from about $j=9$ where $n_{\bar{A} k}$ starts to decrease again. The graphs for $n_{\bar{A} k}$ for various values of $j$ are presented in figure \ref{figure:ngtmu0p01xi500}.

\begin{figure}
 \centering
 \includegraphics[scale=0.2]{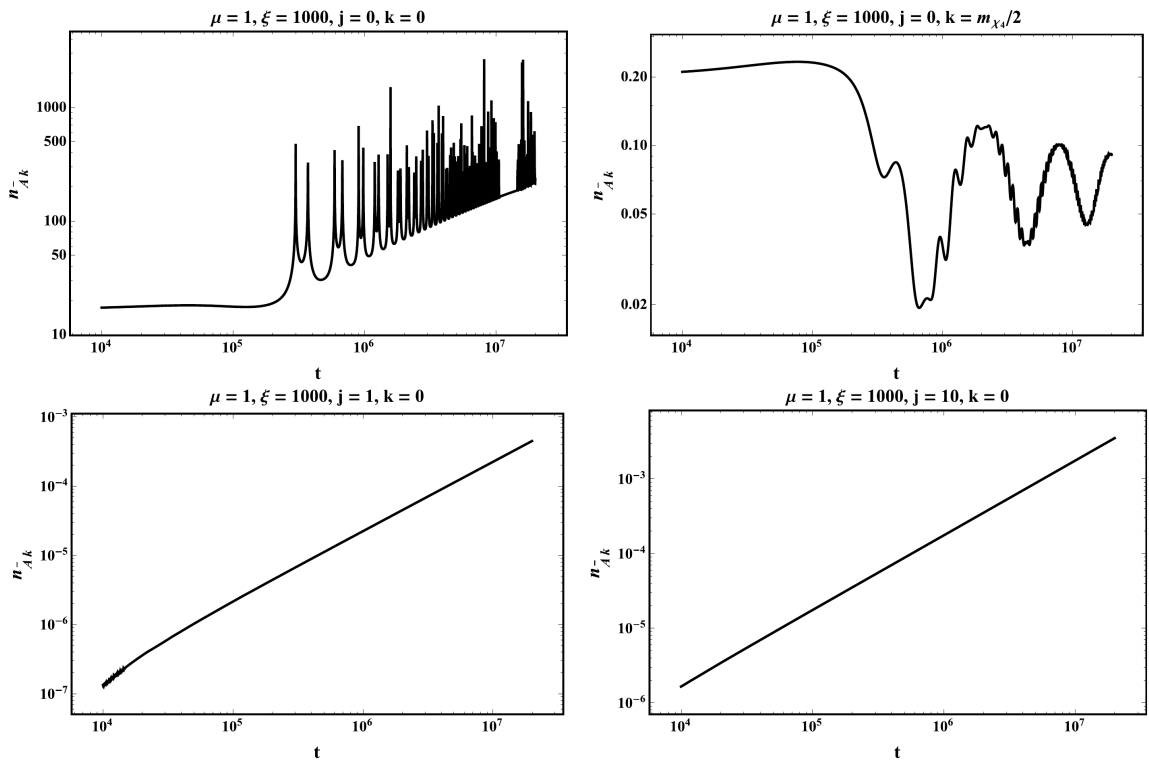}
   \caption{The time evolution of number density of various gauge preheat modes, $n_{\bar{A} k}$, for different $j's$ and $k$'s $n_{\bar{A} k}$ for $\mu = 1~M_{P}$ and $\xi = 1000$.}
 \label{figure:ngtmu1xi1000}
\end{figure}

\item $\mu=0.01~M_{P}$ and $\xi=1520$: In this case the inflaton interpolates between the symmetric and symmetry breaking vacuum, until it settles oscillating around the symmetric vacuum. Since the mass of the gauge modes become zero whenever the inflaton passes through the symmetric vacuum, we expect to see a burst of particle production. Even for $j=0$ and $k=0$, that the mass of the gauge mode is independent of the inflaton, we again see spiky behavior in the number density of the produced massless gauge mode, fig. \ref{figure:ngtmu0p01xi1520}, which happens in a narrow resonance band, since for $k=m_{\chi_2}/2$, the amplitudes of the spikes gets completely suppressed. For $j=1$, the stochastic particle production is more enhanced, but similar to the previous case, the resonance happens in a broad resonance band. Even for $k=4~m_{\chi_4}$, the amplitude of spikes are still quite large. This suggests that $j=1$ gauge field production is quite efficient at the end of $\MM$-flation. With increasing $j$, the amplitude of the produced particles grows substantially, since the inflaton starts oscillating around the symmetric vacuum at which the gauge modes become massless and the production becomes easier.

\begin{figure}
 \centering
 \includegraphics[scale=0.2]{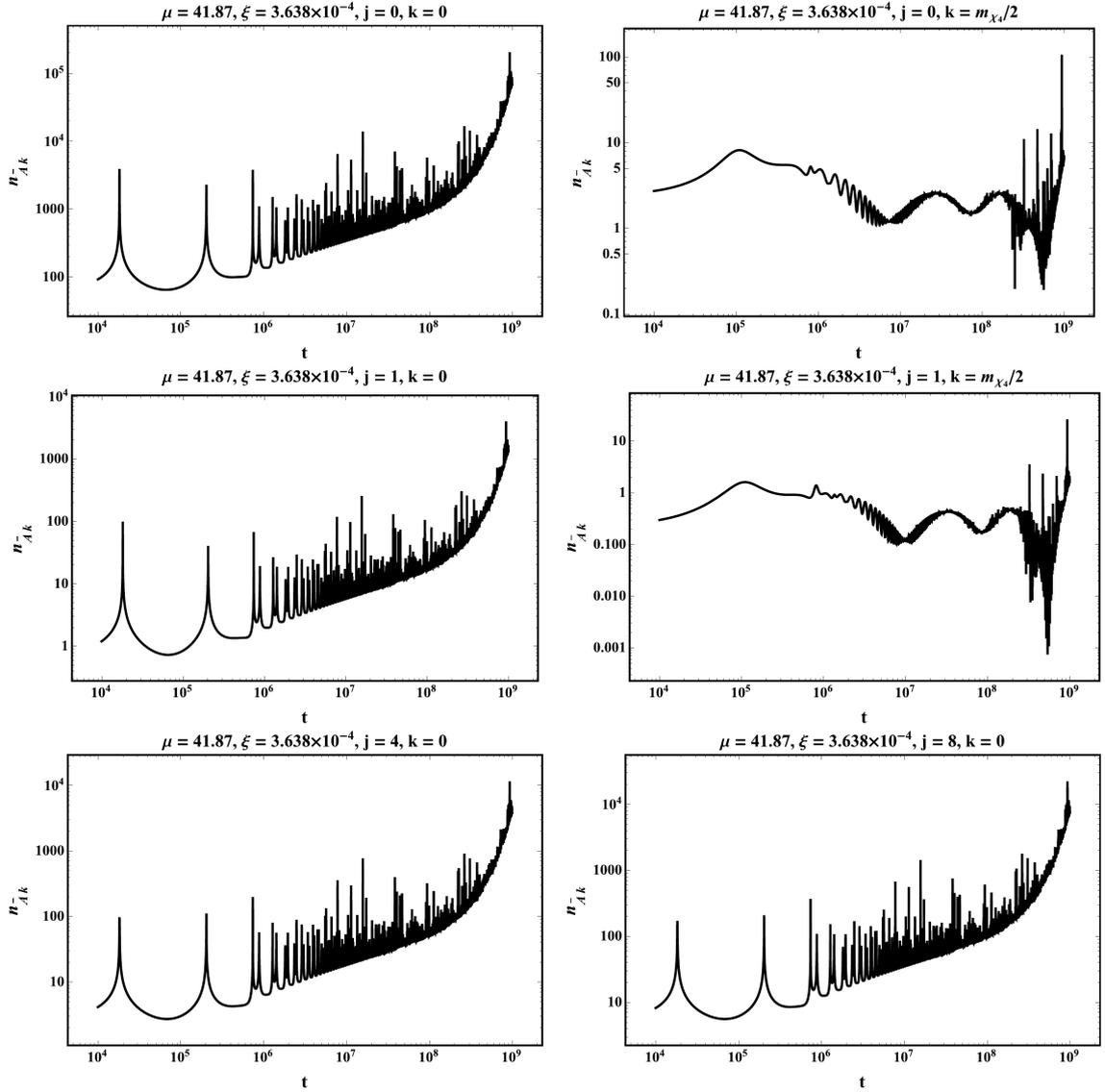}
   \caption{The time evolution of number density for several $j$ gauge modes, $\bar{A}_i$ for $\mu = 41.87~M_{P}$ and $\xi = 3.638\times 10^{-4}$.}
 \label{figure:ngtmu41p87}
\end{figure}

 \item  $\mu=1~M_{P}$ and $\xi=1000$: The symmetry-breaking vacuum in this case takes Planckian value and the non-minimal coupling parameter is large. The inflaton only oscillates around the symmetry-breaking vacuum. Aside from $j=0$ gauge mode, which is produced in a very narrow resonance band, as it is manifest from the figure  \ref{figure:ngtmu1xi1000}, the production of higher gauge modes are all suppressed. This is understood by the fact that the bare mass ({\it i.e.} the inflaton-independent part of the mass, gets larger with $j$ and they become harder to produce, see fig. \ref{figure:ngtmu1xi1000}.

  \item  $\mu=10~M_{P}$ and $\xi=100$: The symmetry-breaking vacuum  in this case, takes super-Planckian values in the Jordan frame and the non-minimal coupling parameter is moderately large. The behavior of number density of the gauge modes as a function of $j$ parameter is similar to the previous case: spiky enhancements for $j=0$ and $k=0$ in a narrow resonance band and lack of particle production for higher $j$'s. We can derive this conclusion that for (super-)Planckian $\mu$'s and $\xi\gg 1$, except for a narrow resonance band around $k=0$ for $j=0$, the number density of produced particles through preheating is suppressed.

  \item  $\mu=41.87~M_{P}$ and $\xi=3.638\times 10^{-4}$: In this case even though $\mu$ is super-Planckian and non-minimal coupling is tiny, aside from a narrow resonance band around $k=0$ for $j=0$, larger values of $j$'s get a similar spiky behavior in the number density of the produced particles, $n_{\bar{A} k}$, which increases with time, please see fig. \ref{figure:ngtmu41p87}.

\end{itemize}

\section{UV cutoff in non-$\mathbb{M}$-flation}
\label{UVcutoff}

Higgs inflation \cite{Bezrukov:2007ep}, with its non-minimal coupling to gravity, although is an appealing scenario that uses the now-known-to-exist ingredient of the standard model, namely the scalar field Higgs, to address the origin of structures, fails to address the issue of the displacements beyond the UV cutoff in the model. In \cite{Burgess:2010zq, Barbon:2009ya, Hertzberg:2010dc}, disregarding the fact that the value of the inflaton during Higgs field during inflation is large the value of the cutoff in both Jordan and Einstein frames are obtained to be $\Lambda\sim \frac{M_P}{\xi}$.  However noting the above missing point, it has been discusses in \cite{Bezrukov:2010jz} that the UV cutoff of the theory in the Jordan frame depends on the background value of the Higgs field during inflation, where $\phi>M_{P}/\sqrt{\xi}$, to be $\Lambda^{J}\simeq \sqrt{\xi}\phi$. Still going to the Einstein frame \cite{Bezrukov:2010jz} demonstrate that the cutoff of the theory remains to be $\Lambda_{E}\simeq M_P$.  Although in both frames such energies are beyond the energy scales involved in the Higgs inflation, one can easily see that in particular in the Einstein frame where the scalar field, $\chi$, is canonical and the gravity takes the Einstein-Hilbert form, the displacements of the  field remains super-Planckian, $\Delta\chi \gg M_P$. This raises the question of robustness of the Higgs inflation to higher dimensional operators and in particular the six-dimensional operators that can resurrect the $\eta$-problem in Higgs inflation.

In non-$\MM$-flation, the presence of non-minimal coupling acts similar to what it does in Higgs inflation, although the non-minimal coupling, $\xi$, here could take much smaller values. As discussed above, with $\xi\sim {\rm few}\times 100$, one can reduce the number of number of D3-branes to values as small as ${\rm few}\times 10$. If we go to Einstein frame, the UV cutoff of the theory, as \cite{Bezrukov:2010jz} have argued for is $M_P$. Now in the Einstein frame the effective canonical $\chi$ field traverses ${\rm few}\times M_P$. For example in the case where $\mu=0.01 M_{P}$ and $\xi=1520$, $\Delta \chi\simeq 5.12 M_{P}$. However, as we mentioned before, with the number of D3 branes needed to reduce the bare coupling $\lambda=8\pi g_{_{S}}\simeq 1$ to $1.6\times 10^{-4}$ needed for CMB observations, $N\simeq 37$, the typical {\it physical} field displacements is then reduces to $\frac{2\Delta\chi}{\sqrt{N (N^2-1)}}\simeq 0.04~M_P$, which is below $M_P$.

Still with the number of species involved during (non)-$\MM$-flation, one should be worried about the reduction of Planck mass \cite{Dvali:2007hz}. In the case of gauged $\MM$-flation, in \cite{Ashoorioon:2011ki}, it was argued that only light modes contribute to lowering the UV cutoff. It was suggested that light would mean only the modes that are lighter than the Hubble parameter during inflation. In \cite{Ashoorioon:2011aa}, we pointed out that gauged $\MM$-flation also enjoys a hierarchical characteristic that can preclude all the modes from contributing to lowering the species UV cutoff. These two papers, we now think, miss a nice property of the gauged model that can better address the issue, and that is how the gauge modes contribute to the running of the Planck mass. We noted above that the gauged (non-)$\MM$-flation has $3N^2-1$ gauge vector modes and $2N^2+1$ degrees of freedom coming from the scalar sector (the $SU(2)$ sector inflaton is included). It has been shown that in presence of matter field fluctuations, the strength of the gravitational interactions is modified. The Planck mass at the energy scale $\mu$ is
\beq
M(\mu)^2=M_P^2-\frac{\mu^2}{12\pi}(N_s+N_{f}-4 N_v)
\eeq
where $N_s$, $N_f$ and $N_v$ are respectively the number of spin zero, spin one half and spin one gauge vector bosons \cite{Larsen:1995ax, Calmet:2008df}. $M_P$ is the value of Planck mass at zero energy. When only scalar species are involved, there is an energy scale at which the Planck mass $M(\mu)$ becomes zero, or respectively when the gravitational coupling becomes infinite. When only scalar species are involved, there is a cutoff scale at which this happens. This is known as the species UV cutoff, described by Dvali  as \cite{Dvali:2007hz},
\beq
\Lambda=\frac{M_P}{\sqrt{N_s}}\,.
\eeq
The same is true with the fermion species, {\it i.e.} they contribute with the similar sign as scalars, and under their effect the Planck mass is lowered from the zero energy Planck mass. However gauge degrees of freedom counteract the effect of scalars and fermions, if there are enough of them in the theory. In particular in gauged (non-)$\MM$-flation, we will have
\beq
M(\mu)^2=M_P^2+\frac{\mu^2}{12\pi}(10 N^2-5)\,,
\eeq
and for $N\geq 1$, this always runs to larger values than $M_P$ as $\mu$ increases. Hence, because of gauge vector modes in (non-)$\MM$-flation, not  only with the multitude of species that exists, the UV cutoff of the theory is not reduced, but enhances as we run from the zero energy to the scale of inflation which is about the GUT scale.

Since (non)-$\MM$-flation is part of a grander supersymmetric scheme in which fermions are also involved, one may get worried that the presence of the fermions startles the nice feature that we obtain considering only the gauge and scalar degrees of freedom. However, one should note that the degrees of freedom of the fermions are $5N^2$ and in their presence, the running will be modified as
\beq
M(\mu)^2=M_P^2+\frac{\mu^2}{12\pi}(5 N^2-5)\,,
\eeq
which again for $N\geq 1$, either does not induce any running in the Planck mass (for $N=1$) or enhances it from its value at zero energy. We conclude that contrary to N-flation and Higgs inflation, (non-)$\MM$-flation is not afflicted with the problem of excursions beyond the UV cutoff.


\section{Conclusions}
\label{section:conclusions}

In this paper, we investigated the consequences of non-minimal coupling to gravity in an inflationary model driven with matrices in an appropriate flux background, which we called non-$\MM$-flation. We suggested two mechanisms that such non-minimal coupling can arise in a string theory setup, either from loop corrections of the species to the graviton-scalar-scalar or from the dependence of the superpotential on the brane position. Such corrections were invoked in the past to cancel the conformal coupling to graviton in brane-antibrane inflation in the KKLT setup. The former can hardly produce the non-minimal coupling $\xi\gtrsim 1$, but in the latter, one can in principle obtain larger values for $\xi$. We showed that with $\xi\sim 10^{-2}$, the predictions of non-$\MM$-flation in the symmetry-breaking region, $\phi>\mu$, becomes compatible with the latest PLANCK data. With non-minimal coupling of $\sim {\rm few}\times 10^2$, the number of D3-branes required in the string theory setup to reduce the bare coupling of order one, can be reduced to $N\lesssim 10^2$.  It also further reduces the amplitude of isocurvature modes at the end of inflation.

It was also shown that the effect of non-minimal coupling can also address the issue of embedded preheating in the model. This is achieved in two ways. One picture that can arise is that because of the non-minimal coupling, the symmetry-breaking vev can take sub-Planckian values and, after the termination of inflation, the inflaton has enough energy to roll over the bump in the potential to interpolate between the symmetry-breaking and symmetric vacua. In this case, the preheating is quite successful as around the symmetric vacuum,  the spectator modes become massless or tachyonic, allowing for a burst of spectator field(s) production to occur, depleting the energy of the inflaton. The other scenario that can mostly occur if the symmetry-breaking vacuum is (super-) Planckian, is that for the light spectator modes, narrow resonance bands can occur and be sustained despite the expansion of the universe. In such scenarios, it is expected that only a fraction of the energy of the inflaton in the $SU(2)$ sector is transferred to the spectator fields. Still since the energy of the inflaton is about the $\sim M _{\rm GUT}^4$, and the reheating temperature has only have to be larger than MeV to satisfy the nucleosynthesis constraints, it is expected that even a small fraction of transfer of energy of the inflaton, will help us in achieving this purpose. We will investigate the amount of energy transfer to the spectator fields in such scenarios  and the consequences of it to the shape of the inflaton potential to future studies. In the investigation of preheating for the gauge modes, we noticed that the $U(1)$ sector of the $U(N)$ gauge fields, which remained massless, produces transient spikes in the spectrum of particles produced. This is emanated from the violation of the adiabaticity condition for these modes. Such produced (electro-)magnetic fields will have a very tiny wavelengths. Although it sounds difficult to relate such produced magnetic fields to the observed intragalactic magnetic fields of nano-Gauss strength, this is something that has to be verified in more details.

The other issue that we pointed out in this paper is the fact that the UV cutoff in the model is not reduced from the low energy Planck mass, despite the existence of multitude of species in the setup. That has to do with the nature of some of these species, arising from the non-Abelian gauge sector in the model. Such degrees of freedom could be shown to be described perturbatively and effectively by several spin one $U(1)$ modes where all, except for one, are massive. These vector modes causing effectively the gravitational constant becoming infinite at energies smaller than the Planck mass at the inflationary energy scale. We argued that the presence of fermions do not disturb this nice feature of the model. We also plan to return to the effect of fermions during  (non-)$\MM$-flation (preheating) in future.


\appendix


\section{Non-$\mathbb{M}$-flation in the limit $\xi\gg1$}
\label{section:xigg1}

Here we investigate the non-$\mathbb{M}$-flation in the regime that the non-minimal coupling between the inflaton field and the gravitational field takes values much larger than unity ($\xi\gg1$). In this limit, the conformal factor (\ref{Omega}) reduces to
\begin{equation}
 \label{Omegaxill1}
 \Omega^{2}=\xi\phi^{2}\,.
\end{equation}
From Eq. (\ref{dchidphi}) the relation between the scalar field in the Jordan and the one in Einstein frames is given to be
\begin{equation}
 \label{phichi}
 \phi=\frac{1}{\sqrt{\xi}}e^{\frac{\chi}{\sqrt{6}}}\,.
\end{equation}
Note that in the previous section, we expressed $\phi$ in terms of $\chi$ by the using of the inverse function that should be evaluated numerically, but in this section we find an analytical expression for this purpose. This analytical expression will simplify our calculations considerably. Using this and eq. (\ref{Vphi}) in eq. (\ref{UchiV0phi}), the potential in the Einstein frame is obtained to be
\begin{equation}
 \label{Uchixigg1}
 U(\chi)=\frac{\lambda_{{\rm eff}}}{4\xi^{2}}\left[1-\mu\sqrt{\xi}e^{-\frac{\chi}{\sqrt{6}}}\right]^{2}\,.
\end{equation}

\begin{figure}
 \centering
 \includegraphics[scale=0.7]{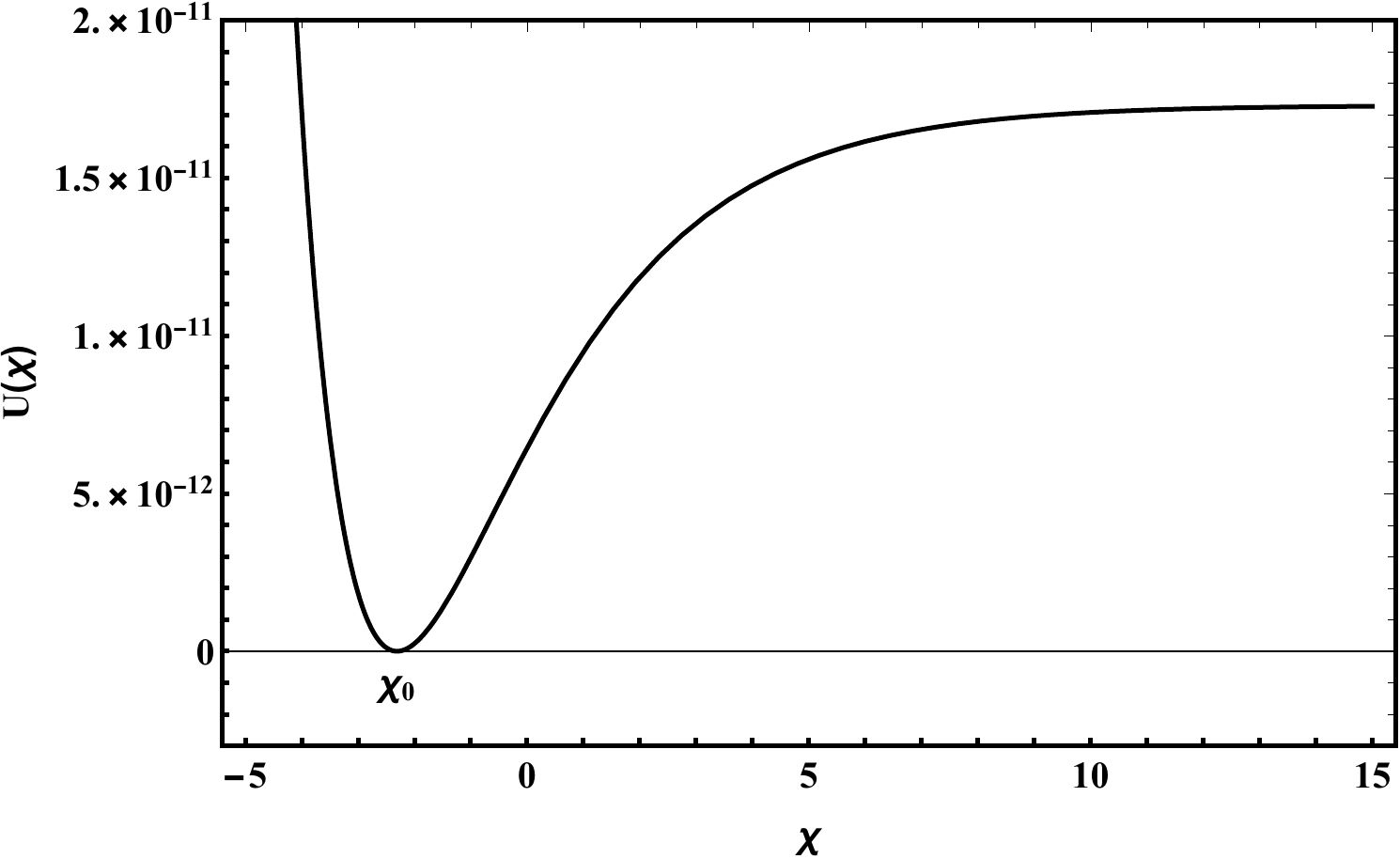}
 \caption{The inflationary potential (\ref{Uchixigg1}) obtained in the limit $\xi \gg 1$.}
 \label{figure:Uchixigg1}
\end{figure}

The diagram of this potential is shown in Fig. \ref{figure:Uchixigg1}. As we see in the figure, this potential, in such a limit, has only one minimum, in contrast with the exact potential (\ref{Uchi}) that has two minima. If we consider the approximation $\xi\gg1$, then we cannot observe one of the minima of the original potential (\ref{Uchi}), whereas in our case where $\mu$ is not necessarily small, it may have phenomenological significances. For instance, the existence of another minimum can give rise to the inflaton rolling over the bump and giving rise to preheating around the symmetric vacuum.

The minimum of the potential (\ref{Uchixigg1}) is placed at
\begin{equation}
 \label{chi0}
 \chi=\sqrt{\frac{3}{2}}M_{P}\ln\left(\frac{\xi\mu^{2}}{M_{P}^{2}}\right)\,.
\end{equation}
Below we will examine the slow-roll inflation on the right hand side of this minimum. With the potential (\ref{Uchixigg1}), the slow-roll parameter (\ref{epsilon}) and (\ref{eta}) are
\begin{align}
 \label{epsilonchixigg1}
 \epsilon & = \frac{1}{3}\left(\frac{1}{\mu\sqrt{\xi}}e^{\frac{\chi}{\sqrt{6}}}-1\right)^{-2},
 \\
 \label{etaxigg1}
 \eta & =-\frac{2}{3}\left(\frac{1}{\mu\sqrt{\xi}}e^{\frac{\chi}{\sqrt{6}}}-1\right)^{-1}
\end{align}
Setting $\epsilon=1$ in Eq. (\ref{epsilonchixigg1}), we find the inflaton at the end of inflation as
\begin{equation}
 \label{chiendxigg1}
 \chi_{{\rm end}}=\sqrt{6}\ln\left[\mu\sqrt{\xi}\left(\frac{\sqrt{3}}{3}+1\right)\right]\,,
\end{equation}
where we have chosen the solution which lies on the right hand side of the minimum (\ref{chi0}). We solve the differential equation (\ref{dchidNe}) by considering the above equation as the initial condition to find the evolution of the inflaton as
\begin{equation}
 \label{chiNexigg1}
 \chi=-\sqrt{6}W_{-1}\left(-\frac{\sqrt{3}+3}{3}e^{-\frac{N_{e}}{3}-\frac{1}{\sqrt{3}}-1}\right)-\sqrt{\frac{2}{3}}N_{e}+\sqrt{6}\ln\left(\frac{\sqrt{3}+3}{3}\mu\sqrt{\xi}\right)-\sqrt{2}\left(\sqrt{3}+1\right).
\end{equation}

In order to calculate the spectrum of the scalar perturbations, we apply Eqs. (\ref{Uchixigg1}) and (\ref{epsilonchixigg1}) in Eq. (\ref{Ps}), and get
\begin{equation}
 \label{Pschixigg1}
 \mathcal{P}_{s}=\frac{\lambda_{{\rm eff}}\mu^{2}}{32\pi^{2}\xi}e^{-\sqrt{\frac{2}{3}}\chi}\left(\frac{1}{\mu\sqrt{\xi}}e^{\frac{\chi}{\sqrt{6}}}-1\right)^{4}.
\end{equation}
In addition, substitution of Eqs. (\ref{epsilonchixigg1}) and (\ref{etaxigg1}) into Eqs. (\ref{ns}) and (\ref{r}) gives the scalar spectral index and tensor-to-scalar ratio as
\begin{align}
 \label{nschixigg1}
 n_{s} & =\frac{\mu^{2}\xi+3e^{\sqrt{\frac{2}{3}}\chi}-8\mu\sqrt{\xi}e^{\frac{\chi}{\sqrt{6}}}}{3\left(e^{\frac{\chi}{\sqrt{6}}}-\mu\sqrt{\xi}\right)^{2}},
 \\
 \label{rchixigg1}
 r & =\frac{16\mu^{2}\xi}{3\left(e^{\frac{\chi}{\sqrt{6}}}-\mu\sqrt{\xi}\right)^{2}}.
\end{align}
The number of e-foldings at the moment of horizon crossing is regarded as the varying parameter, which depends on the scale of inflation and reheating temperature and it usually varies in the domain $50\leq N_{e}\leq60$. For $N_e = 50$ ($60$), the scalar spectral index and tensor-to-scalar ratio are obtained as $n_{_{S}} = 0.9630$ ($0.9690$), and $r = 0.0136$ ($0.0098$), respectively.  An important point is that in the limit $\xi\gg1$, the prediction of the model for $n_{_{S}}$ and $r$ are independent of the parameter $\mu$, so that the plot of the model in this regime is the same in all graphs of Fig. \ref{figure:rnsmu-chi-bigger-chi4}. These observables are also independent of the parameter $\lambda_\mathrm{eff}$, which should be determined by fixing the amplitude of the scalar power spectrum (\ref{Pschixigg1}) at the horizon exit according to the Planck 2018 data \cite{Akrami:2018odb}.


\section{Revising the Higgs inflation Predictions}
\label{section:Higgs}

In the model of Higgs inflation \cite{Bezrukov:2007ep}, the Higgs boson of the standard model of particle physics is regarded as the inflaton. The action of the model in the Jordan frame as the physical frame of the model, is
\begin{equation}
 \label{SJHiggs}
 S_{J}=\int d^{4}x\sqrt{-g}\left[\frac{M_{P}^{2}}{2}\left(1+\frac{\xi}{M_{P}^{2}}\phi^{2}\right)R+\frac{1}{2}\partial_{\mu}\phi\partial^{\mu}\phi-V(\phi)\right].
\end{equation}
Here, $V(\phi)$ is the Higgs potential,
\begin{equation}
 \label{VphiHiggs}
 V(\phi)=\frac{1}{4}\lambda\left(\phi^{2}-\nu^{2}\right)^{2}.
\end{equation}
The parameter $\nu$ is the vacuum expectation value of the Higgs field, and $\lambda$ is the Higgs self-coupling parameter. The Higgs mass has relation with these parameters as $m_{H}=\sqrt{2\lambda}\:\nu$. The expectation value of the Higgs field is given by $\nu=\left(\sqrt{2}G_{F}\right)\approx246\,\textrm{GeV}$, where $G_F$ is the Fermi coupling which is determined with a precision of 0.6 ppm from muon decay measurements \cite{vanRitbergen:1999fi,  Steinhauser:1999bx, Webber:2010zf}. Also, the Higgs mass is measured in the ATLAS \cite{Aad:2012tfa} and CMS \cite{Chatrchyan:2012xdj} experiments as $m_{H}\approx125\,\textrm{GeV}$. Combining this result, the value of the Higgs self-coupling parameter is $ \lambda\approx0.129$.

To study the Higgs inflation model, it is more appropriate to go to the Einstein frame through the conformal transformation (\ref{gtildemunu}). The conformal factor of this transformation is
\begin{equation}
 \label{OmegaHiggs}
 \Omega^{2}=1+\frac{\xi}{M_{P}^{2}}\phi^{2}.
\end{equation}
In the Einstein frame, the action takes the form
\begin{equation}
 \label{SEHiggs}
 S_{E}=\int d^{4}x\sqrt{-\tilde{g}}\left[\frac{M_{P}^{2}}{2}\tilde{R}+\frac{1}{2}\partial_{\mu}\chi\partial^{\mu}\chi-U(\chi)\right].
\end{equation}
Using Eq. (\ref{dchidphi}), the field $\chi$ is expressed in terms of $\phi$ as
\begin{equation}
 \label{chiphiHiggs}
 \chi\equiv f(\phi)=\frac{M_{P}}{\sqrt{\xi}}\left[\sqrt{7}\,\ln\left(\sqrt{7\xi\left(\frac{7\xi\phi^{2}}{M_{P}^{2}}+1\right)}+\frac{7\xi\phi}{M_{P}}\right)-\sqrt{6}\,\tanh^{-1}\left(\frac{\sqrt{6\xi}\,\phi}{\sqrt{7\xi\phi^{2}+M_{P}^{2}}}\right)\right].
\end{equation}
The previous investigations on the Higgs inflation are usually performed by assuming $\sqrt{\xi}\,\phi\gg M_{P}$ (see \cite{Bezrukov:2007ep}). With this assumption, the second term in the above equation can be dropped versus the first one, and the calculations would be similar to what we performed in appendix \ref{section:xigg1}. However it is more precise if we follow the approach of sec. \ref{section:setup} in the study of non-$\MM$-flation. In this approach the full expression of eq. (\ref{chiphiHiggs}) is kept. Of course, in this procedure we cannot find $\phi$ in terms of $\chi$ analytically, but we can employ the inverse function in Mathematica, $\phi=f^{-1}(\chi)$, which can evaluate $\phi$ numerically for a given value of $\chi$. From eq. (\ref{UchiV0phi}) the potential in the Einstein frame will be
\begin{equation}
 \label{UchiHiggs}
 U(\chi)=\frac{\lambda}{4}\left[\nu^{2}-\left(f^{-1}(\chi)\right)^{2}\right]^{2}\left[\frac{\xi\left(f^{-1}(\chi)\right)^{2}}{M_{P}^{2}}+1\right]^{-2}\,.
\end{equation}
The potential has two minima at
\begin{align}
 \label{chi1Higgs}
 \chi_{1} & =f(-\nu),
 \\
 \label{chi3Higgs}
 \chi_{3} & =f(\nu)\,,
\end{align}
and a maximum between these two minima at
\begin{equation}
 \label{chi2Higgs}
 \chi_{2}=\sqrt{\frac{7}{\xi}}\,\ln\left(\sqrt{7\xi}\right)\,.
\end{equation}
The Einstein frame potential which is resulted in the previous studies of the Higgs inflation \cite{Bezrukov:2007ep} has only one minimum, but another minimum is appeared when we consider the full expression in Eq. (\ref{chiphiHiggs}). The other minimum, which appears in this approach, may have notable consequences for the reheating process as we noted above.

Following the same procedure we used in sec. \ref{section:setup}, we estimate the inflationary observables $n_{_{S}}$ and $r$ in the Higgs inflation model. The explicit values of $n_{_{S}}$ and $r$ with $N_e=50$ and $N_e= 60$ are presented in table \ref{table:nsrHiggs}. In the table, we have compared the results of our approach with the approximate method used \cite{Bezrukov:2007ep}. In the inexact approach, the scalar spectral index and the tensor-to-scalar ratio are calculated respectively to be \cite{Bezrukov:2007ep}
\begin{align}
 \label{nsHigss}
 n_{s} & =1-\frac{8\left(4N_{e}+9\right)}{\left(4N_{e}+3\right)^{2}}\,,
 \\
 \label{rHiggs}
 r & =\frac{192}{\left(4N_{e}+3\right)^{2}}\,.
\end{align}
From the Table \ref{table:nsrHiggs}, we see that the results of the exact approach is very close to the approximate method used by \cite{Bezrukov:2007ep}, but there exists a slight difference which can be important in the light of future precise measurements of the CMB \cite{Abazajian:2016yjj}.

\begin{table}[t]
  \centering
  \caption{Values of the scalar spectral index $n_{_{S}}$ and tensor-to-scalar ratio $r$ for the Higgs inflation model in our accurate approach in comparison with the previous approximate approach \cite{Bezrukov:2007ep}.}
  \scalebox{1.0}{
  \begin{tabular}{lccc}
    \hline
    \hline
    & $\qquad N_{e}\qquad$ & $\qquad$ Exact approach $\qquad$ & $\qquad$ Previous approach $\qquad$
    \tabularnewline
    \hline
    \multirow{2}{*}{$n_{s}$ $\qquad$} & \multicolumn{1}{c}{$50$} & $0.9616$ & $0.9594$\tabularnewline
    & $60$ & $0.9678$ & $0.9663$\tabularnewline
    \hline
    \multirow{2}{*}{$r$ $\qquad$} & $50$ & $0.0042$ & $0.0047$\tabularnewline
    & $60$ & $0.0030$ & $0.0033$\tabularnewline
    \hline
  \end{tabular}
  }
  \label{table:nsrHiggs}
\end{table}

\section*{Acknowledgement}

We are thankful to Mustafa A. Amin, Ram Brustein, Andrei Linde, Liam McAllister, M. Shaposhnikov, Shinji Mukohyama, and Gary Shiu for helpful discussions. In particular we are thankful to M. M. Sheikh-Jabbari for many illuminating inputs. A. A. is grateful to the theoretical division of CERN, where part of this project was fulfilled.

\bibliographystyle{JHEP}
\bibliography{bibtex}

\end{document}